\documentclass[aps, pra, superscriptaddress, reprint]{revtex4-2}

\usepackage{hyperref}
\usepackage{physics}
\usepackage{xcolor}
\usepackage{amssymb}
\usepackage{mathtools}
\usepackage{bbm}
\usepackage{bm}
\usepackage{mathrsfs}
\usepackage{verbatim}
\usepackage{dsfont}

\makeatletter
\DeclareRobustCommand{\cev}[1]{%
  {\mathpalette\do@cev{#1}}%
}
\newcommand{\do@cev}[2]{%
  \vbox{\offinterlineskip
    \sbox\z@{$\m@th#1 x$}%
    \ialign{##\cr
      \hidewidth\reflectbox{$\m@th#1\vec{}\mkern4mu$}\hidewidth\cr
      \noalign{\kern-\ht\z@}
      $\m@th#1#2$\cr
    }%
  }%
}
\makeatother

\begin{document}

\title{Unifying quantum stochastic methods using Wick's theorem on the Keldysh contour}

\date{\today}

\author{Vasco Cavina}\email{vasco.cavina@sns.it}
\affiliation{Scuola Normale Superiore, 56126 Pisa, Italy}

\author{Antonio D'Abbruzzo}
\affiliation{Scuola Normale Superiore, 56126 Pisa, Italy}

\author{Vittorio Giovannetti}
\affiliation{Scuola Normale Superiore, NEST, and Istituto Nanoscienze-CNR, 56126 Pisa, Italy}

\begin{abstract}

We present a novel method, based on the Keldysh formalism, for deriving stochastic master equations that describe the non-Markovian dynamics of a quantum system coupled to a Gaussian environment. This approach yields a compact expression for the system's propagator, which we show to be equivalent to existing formulations, such as the stochastic von Neumann equation (SVNE). A key advantage of our method is its generality: it can be extended to describe any open-system evolution defined on a suitable ordering contour. As a result, we adapt it to derive generalized versions of the SVNE that account for initial system-environment correlations, as well as new stochastic equations that incorporate information about the statistics of energy flows in the environment.
The insights offered by our technique further allow us to examine the nature of the noise processes appearing in the SVNE. We prove that its solution can be expressed in terms of a single physical noise, without any loss of information. Finally, we propose a semiclassical scenario in which this noise can be interpreted as arising from an initial measurement process on the environment.

\end{abstract}

\maketitle


\section{\label{sec:intro}
    Introduction
}

One of the important problems of contemporary quantum physics is finding convenient ways to describe the intricate dynamics of open quantum systems~\cite{breuer2007book}.
In the Markovian regime, where the excitations injected into the environment are not reflected back into the system, the well-known Lindblad-Gorini-Kossakowski-Sudarshan theorem~\cite{lindblad1976classic,gorini1976classic} provides a full characterization of the dynamical equation, which has been thoroughly studied (even though the question of its microscopic derivation is still open in certain regimes, see, e.g., Refs.~\cite{nathan2020ule,mozgunov2020positive,mcCauley2020lindblad,dabbruzzo2023reg,dabbruzzo2024choi} and references therein).
In the non-Markovian scenario, the situation is much less clear and a great variety of approaches has appeared in the literature during the years~\cite{deVega2017review}.

At the same time, nonequilibrium quantum field theory has been able to provide important results, e.g., in cosmology, high energy physics, and condensed matter physics~\cite{stefanucci2013book,kamenev2011book}.
The Keldysh contour formalism~\cite{schwinger1961classic,keldysh1965classic}, in particular, has been proved to be highly efficient for treating perturbative diagrammatic calculations for many-body systems and their classical limit.

Although several works have addressed problems in thermodynamics and open quantum systems from the perspective of the Keldysh contour~\cite{sieberer2016driven,muller2017diagrams,funo2018pathint,reimer2019density,mcDonald2023thirdquant,cavina2023convenient,osorio2024spinboson,cavina2024quantum}, we believe that the Keldysh formalism still holds substantial untapped potential for advancing research in these fields.
In this paper, we exploit this potential by presenting a series of results obtained by formulating the problem of non-Markovian quantum dynamics on the Keldysh contour, with a particular focus on stochastic unravelings of the quantum dynamics.

In Sec.~\ref{sec:keldysh}, we introduce the formalism and detail how the Hamiltonian evolution of an open quantum system can be reformulated using the Keldysh contour, highlighting the advantages this approach offers when the environment is Gaussian and linearly coupled to the system. In this case, Wick's theorem allows us to derive a compact expression for the system's dynamics, akin in spirit to the Feynman-Vernon influence functional \cite{feynman1963influence}.

Up to this point, we have not introduced any stochastic description, focusing instead on deriving compact deterministic expressions for the reduced system evolution. By contrast, in Sec.~\ref{sec:svne} we proceed by representing the environment correlation function using random fields.
In this way we obtain an exact stochastic non-Markovian equation for the dynamics of the system
in which the noises are defined on the Keldysh contour, that turns out to be equivalent to other stochastic approaches \cite{diosi1998diffusion,stockburger2001diffusion} and in particular to the the so-called stochastic von Neumann equation (SVNE)~\cite{kubo1969stochastic,stockburger2002exact,stockburger2004simulating,tanimura2006review,tilloy2017unraveling}.
In this context, a Keldysh-like contour was already being employed in Ref.~\cite{stockburger2002exact} but with arguably more involved path-integration methods.
The main distinction between the approach of this section and the previous attempts is methodological, since we directly apply Wick's theorem with respect to the Keldysh ordering~\cite{stefanucci2013book}, contrary to existing approaches, based on path integration \cite{feynman1963influence,hedegard1987fermion} or superoperator representations~\cite{diosi2014gaussian,cirio2022influence}.

We fully exploit our methodology in Sec.~\ref{sec:extensions}, where we derive two fundamental extensions of the theory. First, we study the case in which the system and environment share initial equilibrium correlations \cite{grabert1988quantum}, and show that these can be elegantly incorporated into the formalism. Second, we apply our technique to quantum thermodynamics, deriving a generalization of the SVNE to study heat statistics within the two-point measurement scheme. To the best of our knowledge, this equation is new, and we discuss its potential relevance for the community.

From a more formal and theoretical perspective, our method enables us to make insightful statements about the nature of the noise terms appearing in the SVNE.
Specifically, in Sec.~\ref{sec:reduction} we show, using a ``partial'' Wick's theorem, that it is possible to write the solution to the SVNE in terms of a single real noise with a positive semidefinite autocorrelation function.
Moreover, in Sec.~\ref{sec:measurements} we show how the SVNE is influenced by a measurement on the environment performed before it starts interacting with the system.
We find that if the environment satisfies appropriate semiclassical conditions, it is possible to interpret the noise with a positive semidefinite autocorrelation appearing in the SVNE in terms of the outcomes of a single heterodyne initial measurement.
This provides additional insights into the open problem of determining whether non-Markovian stochastic equations admit a general measurement interpretation~\cite{gambetta2002measurement,diosi2008retarded,wiseman2008pure,megier2020measurement, kronke2012non}.
Finally, in Sec.~\ref{sec:conclusions} we draw our conclusions.


\section{\label{sec:keldysh}Deterministic quantum dynamics using the Keldysh contour
}

\subsection{Quantum dynamics and the Keldysh contour}

Let us consider a quantum system $S$ whose state is described at time $t$ by a density operator $\rho(t)$.
Assuming the state $\rho(0)$ at an initial time $t = 0$ to be known,
we have
\begin{equation} \label{eq:vNeu_sol}
    \rho(t) = U(t,0) \rho(0) U^\dagger(t,0),
\end{equation}
where $ U(t,0) $ is the time-ordered exponential of the (possibly time-dependent) Hamiltonian $H(t)$ that generates the dynamics of the system.
After introducing the time-ordering operator $\cev{\mathcal{T}}$ and the anti time-ordering operator $\vec{\mathcal{T}}$, which respectively rearrange their arguments by putting later-time operators to the left and right, we can expand Eq. \eqref{eq:vNeu_sol} as
\begin{multline} \label{eq:vNeu_sol_expanded}
    \rho(t) = \sum_{n,m=0}^\infty \frac{(-i)^{n+m}}{n! m!} \int_0^t d\tau_1 \ldots d\tau_n \int_{t}^0 d\tau'_1 \ldots d\tau'_m \\
    \times \cev{\mathcal{T}} \qty{ H(\tau_1) \ldots H(\tau_n) } \rho(0) \vec{\mathcal{T}} \qty{ H(\tau'_1) \ldots H(\tau'_m) }.
\end{multline}

\begin{figure}
    \includegraphics[scale=0.73]{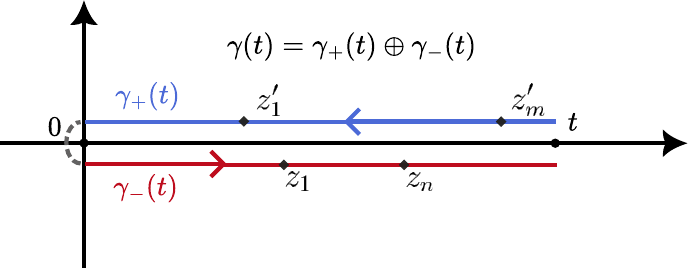}
    \caption{
        Schematic depiction of the Keldysh contour used in this paper.
        The contour $\gamma(t)$ is composed by a backward branch $\gamma_+(t)$ running from $t$ to $0$ and a forward branch $\gamma_-(t)$ running from $0$ to $t$, in this order.
        For the sake of clarity, we added some of the integration variables appearing in Eq. \eqref{eq:againdens}. The variables $z_1 = \tau_{1 -},z_n = \tau_{n -}$ and $z_1'= \tau_{1 +}', z_m' = \tau_{m  +}'$ belong to the $\gamma_-$ and $\gamma_+$ branches, respectively, since they originally represent contributions coming from the time-ordered and anti time-ordered parts of Eq. \eqref{eq:vNeu_sol_expanded}.}
    \label{fig:keldysh}
\end{figure}

In the above equation, $\int_0^{t} d\tau_1 \dots d\tau_n$ is used as a shorthand for $\int_0^{t} d\tau_1 \dots \int_0^{t} d\tau_n$, omitting repeated integration symbols for simplicity. Notice that we also set $\hbar = 1$.
A further simplification of the notation can be achieved by introducing the {\it Keldysh contour},
a convenient choice of integration domain that allows one to write Eq.~\eqref{eq:vNeu_sol_expanded} as a single ordered product ~\cite{stefanucci2013book}.
We begin by denoting the domain of $\tau_1, ..., \tau_n$ in Eq. \eqref{eq:vNeu_sol_expanded} as $\gamma_-(t)$ and distinguishing it from the domain of $\tau_1', ..., \tau_n'$, which we denote as $\gamma_+(t)$.
Both domains are copies of $[0,t]$ and, for ordering purposes, is convenient to regard $\gamma_-(t)$ to be positively oriented when going from $0$ to $t$ and $\gamma_+(t)$ to be positively oriented when going from $t$ to $0$.
The Keldysh contour $\gamma(t)$ is obtained by adjoining $\gamma_+(t)$ and $\gamma_-(t)$, in that order (see Fig.~\ref{fig:keldysh}) \footnote{Note that this is the opposite of what is commonly done in many-body physics~\cite{stefanucci2013book,kamenev2011book}, due to the fact that here we are interested in the evolution of the density operator instead of the system observables.}.
Functions defined on $\gamma_-(t)$ and $\gamma_+(t)$, such as the Hamiltonian in Eq. \eqref{eq:vNeu_sol_expanded}, can be promoted to functions on $\gamma(t)$. If we introduce the notation $z = \tau_\pm$ to indicate a point on $\gamma(t)$ that lies at value $\tau$ on $\gamma_\pm(t)$, we define $H(\tau_\pm) \coloneqq H(\tau)$.
Eq.~\eqref{eq:vNeu_sol_expanded} can now be written as
\begin{multline}
    \rho(t) = \!\!\! \sum_{n,m=0}^\infty \frac{(-i)^{n+m}}{n! m!} \int_{\gamma_-(t)} \!\! dz_1 \ldots dz_n \int_{\gamma_+(t)} \!\! dz'_1 \ldots dz'_m \\
    \times \cev{\mathcal{T}} \qty{ H(z_1) \ldots H(z_n) \rho(0) H(z'_1) \ldots H(z'_m) }, \label{eq:againdens}
\end{multline}
where we promoted the time-ordering operator  $\cev{\mathcal{T}}$ to a contour-ordering on $\gamma(t)$ such that its arguments are rearranged by putting later contour-time operators to the left.
It is clear that this acts as the standard time ordering when restricted to $\gamma_-(t)$ and acts as the anti-time ordering when restricted to $\gamma_+(t)$, since the latter goes backward in time.
Regrouping exponentials
\begin{equation}  \label{eq:twoprel}
    \rho(t) = \cev{\mathcal{T}} \qty{ e^{-i \int_{\gamma_-(t)} H(z)dz} \rho(0) e^{-i \int_{\gamma_+(t)} H(z)dz} }.
\end{equation}
If we imagine $\rho(0)$ to be acted upon by the ordering, $\cev{\mathcal{T}}$ places all operators on $\gamma_{+}(t)$ to the right of $\rho(0)$ and all the operators on $\gamma_{-}(t)$ to the left.
Using this property and the fact that operators commute inside the ordering, we have
\begin{multline} \label{eq:vNeu_sol_keldysh}
    \rho(t) = \cev{\mathcal{T}} \qty{ e^{ -i \int_{\gamma(t)} H(z)dz } \rho(0) } \\
    = \sum_{n=0}^\infty \frac{(-i)^n}{n!} \! \int_{\gamma(t)} \!\! dz_1 \ldots dz_n \cev{\mathcal{T}} \qty{ H(z_1) \ldots H(z_n) \rho(0) }.
\end{multline}
This is the main expression from which our discussion starts and should be compared with the arguably more involved one in Eq.~\eqref{eq:vNeu_sol_expanded}.
Let us now assume that our quantum system $S$ is coupled to an external environment $E$, so that the total Hamiltonian of the combined system is of the form
\begin{subequations} \label{eq:totH}
    \begin{gather}
        H = H_0 + V, \\
        H_0 = H_S \otimes \mathbbm{1}_E + \mathbbm{1}_S \otimes H_E, \label{eq:H0} \\
        V = \sum_\alpha A_\alpha \otimes B_\alpha,
        \label{eq:interaction}
    \end{gather}
\end{subequations}
with $H_0$ being the free part and with $V$ being a coupling contribution, written in terms of the Hermitian system operators $A_\alpha$ and the Hermitian environment operators $B_\alpha$~\cite{breuer2007book}.
It is convenient to move to the interaction picture, where the system-environment state $\rho_{SE}(t)$ gets represented as $\varrho_{SE}(t) \coloneqq e^{iH_0 t} \rho_{SE}(t) e^{-iH_0 t}$, and
\begin{equation} \label{eq:VonNeu}
    \frac{d\varrho_{SE}(t)}{dt} = -i [V(t), \varrho_{SE}(t)],
\end{equation}
where $V(t) \coloneqq e^{iH_0 t} V e^{-iH_0 t}$.
According to Eq.~\eqref{eq:vNeu_sol_keldysh}, this is formally solved by
\begin{equation} \label{eq:keldexp}
    \varrho_{SE}(t) = \cev{\mathcal{T}} \qty{ \exp[-i \int_{\gamma(t)} V(z)dz] \varrho_{SE}(0) },
\end{equation}
where $V(z)$ is the trivial generalization of $V(t)$ on the contour.
However, we are interested in the dynamics of the system $S$ alone, which is described by the interaction-picture density operator $\varrho(t) = \Tr_E[\varrho_{SE}(t)]$.
If we assume that system and environment are initially uncorrelated, we have
\begin{equation} \label{eq:uncorrelated}
    \varrho_{SE}(0) = \varrho(0) \otimes \Omega,
\end{equation}
where $\Omega$ is the initial state of the environment.
Then, using also Eq.~\eqref{eq:interaction}, it is immediate to take the partial trace over the environment in Eq.~\eqref{eq:keldexp} to obtain
\begin{multline} \label{eq:ordenv}
    \varrho(t) = \sum_{n=0}^\infty \frac{(-i)^n}{n!} \sum_{\alpha_1,\ldots,\alpha_n} \int_{\gamma(t)} dz_1 \ldots dz_n \\
    \times \Tr[ \cev{\mathcal{T}} \qty{ B_1 \ldots B_n \Omega } ]
    \cev{\mathcal{T}} \qty{ A_1 \ldots A_n \varrho(0) },
\end{multline}
where for compactness we introduced the abbreviations $A_j \coloneqq A_{\alpha_j}(z_j)$ and $B_j \coloneqq B_{\alpha_j}(z_j)$ for $j=1,\ldots,n$, with
\begin{align} \notag
    A_{\alpha}(\tau_\pm) &\coloneqq e^{i H_S \tau} A_{\alpha} e^{-i H_S \tau}, \\
    B_{\alpha}(\tau_\pm) &\coloneqq e^{i H_E \tau} B_{\alpha} e^{-i H_E \tau}.
\end{align}
%


\subsection{\label{sec:gaussian}
    Gaussian environment hypothesis
}

The expression~\eqref{eq:ordenv} for $\varrho(t)$ can be simplified if we assume that the environment $E$ is composed by a collection of independent bosonic modes, such that $H_E$ is quadratic in the corresponding ladder operators~\cite{serafini2017quantum}.
In addition, we require $\Omega$ to be quadratic and the coupling operators $B_\alpha$ to be linear in the ladder operators.
These requirements, which we will collectively identify as ``Gaussian environment hypothesis'', are common to many models across quantum optics and open quantum systems \cite{gardiner2004quantum, breuer2007book} (including paradigmatic frameworks such as the Caldeira-Leggett model \cite{caldeira1983path} and the Dicke model \cite{hepp1973superradiant}).
These requirements imply the validity of Wick's theorem for the environmental correlation functions~\cite{stefanucci2013book,ferialdi2021wick}.
Suppose first that we have
\begin{equation} \label{eq:stability}
    \Tr[B_\alpha(z) \Omega] = 0.
\end{equation}
This is usually known as ``stability condition" in the context of microscopic derivations of quantum master equations~\cite{breuer2007book}.
Then, Wick's theorem tells us that correlation functions with an odd number of operators vanish,
\begin{equation}
    \Tr[ \cev{\mathcal{T}} \qty{ B_1 \ldots B_{2m+1} \Omega } ] = 0,
\end{equation}
while
\begin{multline} \label{eq:Wick1}
    \Tr[ \cev{\mathcal{T}} \qty{ B_1 \ldots B_{2m} \Omega } ] \\
    = \frac{1}{m! 2^m} \sum_{\sigma \in \mathfrak{S}_{2m}} \prod_{j=1}^m C_{\sigma(2j-1), \sigma(2j)},
\end{multline}
where $\mathfrak{S}_{2m}$ is the set of permutations of $\{1,\ldots,2m\}$, and
\begin{equation}  \label{eq:cGF}
    C_{i,j} \equiv C_{\alpha_i,\alpha_j}(z_i,z_j) \coloneqq \Tr[ \cev{\mathcal{T}} \qty{ B_{\alpha_i}(z_i) B_{\alpha_j}(z_j) \Omega } ]
\end{equation}
is the contour Green's function (GF)~\cite{stefanucci2013book,cavina2023convenient} of the environment.
The arguments $z_i,z_j$ in the contour GF \eqref{eq:cGF} are defined on the Keldysh contour in Fig. \ref{fig:keldysh};
the connection between the contour GF and the physical-time GF
\begin{equation} \label{eq:physcorr}
    c_{\alpha\beta}(\tau_1, \tau_2) \coloneqq \Tr[ B_\alpha(\tau_1) B_\beta(\tau_2) \Omega ]
\end{equation}
can be obtained by choosing the arguments $z_i,z_j$ to assume specific values $\tau_{1 \pm}, \tau_{2 \pm}$ on the two branches:
\begin{align} \label{eq:components}
    C_{\alpha,\beta}&(\tau_{1-}, \tau_{2-}) \notag \\
    & = \theta(\tau_1 - \tau_2) c_{\alpha\beta}(\tau_1, \tau_2) + \theta(\tau_2 - \tau_1) c_{\beta\alpha}(\tau_2, \tau_1), \notag \\
    C_{\alpha,\beta}&(\tau_{1+}, \tau_{2+}) \notag \\
    & = \theta(\tau_1 - \tau_2) c_{\beta \alpha}(\tau_2, \tau_1) + \theta(\tau_2 - \tau_1) c_{\alpha\beta}(\tau_1, \tau_2), \notag \\
    & C_{\alpha,\beta}(\tau_{1+}, \tau_{2-}) = c_{\alpha\beta}(\tau_1, \tau_2), \notag \\
    & C_{\alpha,\beta}(\tau_{1-}, \tau_{2+}) = c_{\beta\alpha}(\tau_2, \tau_1).
\end{align}
In the literature on the Keldysh formalism, these functions are also known, respectively, as time-ordered, anti-time-ordered, greater and lesser components of the Green's function~\cite{stefanucci2013book,kamenev2011book}.
Note that $c_{\beta\alpha}(\tau_2, \tau_1) = c^*_{\alpha\beta}(\tau_1, \tau_2)$, as can be seen from the definition in Eq. \eqref{eq:physcorr}.

We now replace Eq.~\eqref{eq:Wick1} inside Eq.~\eqref{eq:ordenv} with $n = 2m$ in order to simplify the latter.
When doing so, we see that all permutations appearing in Eq. \eqref{eq:Wick1} contribute equally to the integral, as can be seen with an appropriate change of variables.
Thus, since $\mathfrak{S}_{2m}$ has cardinality $(2m)!$, we end up with
\begin{multline} \label{eq:gaussianprep}
    \varrho(t) = \sum_{m=0}^\infty \frac{(-1)^m}{m! 2^m} \sum_{\alpha_1,\ldots,\alpha_{2m}} \int_{\gamma(t)} dz_1 \ldots dz_{2m} \\
    \times C_{1,2} \ldots C_{2m-1,2m} \cev{\mathcal{T}} \qty{ A_1 \ldots A_{2m} \varrho(0) },
\end{multline}
which can be rearranged as follows:
\begin{multline} \label{eq:gaussian_exact}
    \varrho(t) = \cev{\mathcal{T}} \bigg\{ \exp \bigg[ - \frac{1}{2} \sum_{\alpha,\beta} \iint_{\gamma(t)} dz_1 dz_2 \\
    C_{\alpha,\beta}(z_1,z_2) A_\alpha(z_1) A_\beta(z_2) \bigg] \varrho(0) \bigg\}.
\end{multline}
Under the previously stated assumptions, this is the most general expression for describing the dynamics of an open system in contact with a Gaussian environment.
Compared to other fully general expressions, such as the one derived in Ref.\cite{diosi2014gaussian}, Eq.\eqref{eq:gaussian_exact} achieves greater compactness due to the use of the Keldysh contour.
In Appendix~\ref{app:gaussian_physical} we show that indeed we recover the result of Ref.~\cite{diosi2014gaussian} once we express Eq.~\eqref{eq:gaussian_exact} in physical-time representation.

For the sake of generality, we extend the result to cases where the stability condition is not satisfied, i.e., when
\begin{equation}
    \Tr[B_\alpha(z) \Omega] \eqqcolon E_\alpha(z) \neq 0.
\end{equation}
This scenario can be treated by introducing the shifted bath operator
\begin{equation}
    \widetilde{B}_\alpha(z) \coloneqq B_\alpha(z) - E_\alpha(z),
\end{equation}
which obviously satisfies the stability condition.
Substituting into Eq.~\eqref{eq:ordenv} and applying Wick's theorem on the product of the shifted operators, one finds
\begin{multline} \label{eq:gaussian_exact_shifted}
    \varrho(t) = \cev{\mathcal{T}} \bigg\{ \exp\bigg[
    -i \sum_\alpha \int_{\gamma(t)} dz \, E_\alpha(z) A_\alpha(z) \\
    -\frac{1}{2} \sum_{\alpha,\beta}
 \iint_{\gamma(t)} dz_1 dz_2 C_{\alpha,\beta}(z_1,z_2) A_\alpha(z_1) A_\beta(z_2)
    \bigg] \varrho(0) \bigg\},
\end{multline}
where this time
\begin{multline} \label{eq:shifted_green}
    C_{\alpha,\beta}(z_1,z_2) = \Tr[ \cev{\mathcal{T}} \{ \widetilde{B}_\alpha(z_1) \widetilde{B}_\beta(z_2) \Omega(0) \} ] \\
    = \Tr[ \cev{\mathcal{T}} \{ B_\alpha(z_1) B_\beta(z_2) \Omega(0) \} ] - E_\alpha(z_1) E_\beta(z_2).
\end{multline}
Eqs.~\eqref{eq:gaussian_exact_shifted} and~\eqref{eq:shifted_green} will prove highly useful in the concluding Sec.~\ref{sec:measurements} of the paper, where we investigate the impact of a weak measurement of the environment on the system's dynamics.
As we will see, the post-measurement states of the environment generally do not satisfy the stability condition.


\section{\label{sec:svne}Stochastic quantum dynamics using the Keldysh contour
}

Even though Eqs.~\eqref{eq:gaussian_exact} and~\eqref{eq:gaussian_exact_shifted} are exact, they are not directly useful in order to solve the dynamics in practical situations,
and this is an important reason why one is typically interested in finding a master equation for $\varrho(t)$ which can then be solved, analytically or numerically.
The master equation corresponding to Eq.~\eqref{eq:gaussian_exact} has been derived in Ref.~\cite{ferialdi2016master} in physical-time representation (see also Appendix~\ref{app:gaussian_physical}).
However, in the general case, the coefficients of the master equation in \cite{ferialdi2016master} turn out to be extraordinarily complicated functions of the microscopic parameters.
A different approach, called stochastic decoupling~\cite{yan2018stochastic}, consists instead in introducing a stochastic operator $R(t)$, whose evolution depends on some stochastic noise variables, such that we recover $\varrho(t)$ after mediating over the noises: $\varrho(t) = \mathbb{E}[R(t)]$.
The idea is that the master equation for $R(t)$ is more easily solved than that for $\varrho(t)$, since the complicated environment effect is encapsulated into a noise that can be numerically simulated.

There are various ways to perform the stochastic decoupling: we will start here with an approach that harmonizes well with our Keldysh contour formalism.
Let us start with Eq.~\eqref{eq:gaussian_exact}, where all the information about the environment is contained into the contour GF.
The idea is to introduce a set of zero-mean Gaussian noises $\xi_\alpha(z)$ on the Keldysh contour, such that
\begin{equation} \label{eq:corranz}
    \mathbb{E}[ \xi_\alpha(z_1) \xi_\beta(z_2) ] = C_{\alpha,\beta}(z_1,z_2).
\end{equation}
Again, let us use the notation $\xi_j \coloneqq \xi_{\alpha_j}(z_j)$.
A scalar form of the Wick's theorem can be written:
\begin{equation} \label{eq:isserlis}
    \mathbb{E}[\xi_1 \ldots \xi_{2m}] = \frac{1}{m! 2^m} \sum_{\sigma \in \mathfrak{S}_{2m}} \prod_{j=1}^m \mathbb{E}[ \xi_{\sigma(2j-1)} \xi_{\sigma(2j)} ].
\end{equation}
We can now start from Eq.~\eqref{eq:gaussian_exact} and perform the steps described in Sec. \ref{sec:gaussian} backwards to arrive at
\begin{multline} \label{eq:dynnewnoise}
    \varrho(t) = \sum_{n=0}^\infty \frac{(-i)^n}{n!} \sum_{\alpha_1,\ldots,\alpha_n} \int_{\gamma(t)} dz_1 \ldots dz_n \\
    \times \mathbb{E}[ \xi_1 \ldots \xi_n ] \cev{\mathcal{T}} \qty{ A_1 \ldots A_n \varrho(0) }.
\end{multline}
If we use the prescription $R(0) = \varrho(0)$, we can now remove the $\mathbb{E}$ sign and recompact the exponential, after substituting $\varrho(t)$ with $R(t)$:
\begin{equation} \label{eq:R_solution}
    R(t) = \cev{\mathcal{T}} \qty{ \exp[ -i \sum_\alpha \int_{\gamma(t)} \xi_\alpha(z) A_\alpha(z) dz ] R(0) }.
\end{equation}
 With the above definition we have $\varrho(t) = \mathbb{E}[R(t)]$ as desired, since Eqs. \eqref{eq:corranz}, \eqref{eq:isserlis}
ensure that when taking averages of Eq. \eqref{eq:dynnewnoise} we obtain Eq. \eqref{eq:gaussianprep}.

Note the similarity between Eq. \eqref{eq:keldexp} and Eq. \eqref{eq:R_solution}, and the fact that Eq. \eqref{eq:R_solution} is written in terms of a (stochastic) system operator only.
Another fundamental difference is that the operator $\xi_\alpha(z) A_\alpha(z)$ is not necessarily the same operator on the two branches of the Keldysh contour, differently from $V(z)$.
On the contrary, in the general case it is necessary for $\xi_\alpha(z)$ to behave differently on the two branches in order to reproduce the branch-dependence of Eq. \eqref{eq:gaussian_exact}.
This asymmetry between the forward and backward branches is fundamental in order to describe effects typically associated with the presence of an external environment, such as dissipation~\cite{kamenev2011book}.

Let us split the contributions of the forward and backward branches by distinguishing the noise valued on $\gamma_-(t)$ from the one defined on $\gamma_+(t)$, i.e., by explicitly introducing $\xi_\alpha(\tau_{\pm})$ inside Eq.~\eqref{eq:R_solution}.
We can now use that the contour ordering in Eq.~\eqref{eq:R_solution} places all the terms containing $\xi_\alpha(\tau_-)$ to the left of $R(0)$, while the ones containing $\xi_\alpha(\tau_+)$ to the right of $R(0)$, obtaining
\begin{multline} \label{eq:componentsR}
    R(t) = \cev{\mathcal{T}} \bigg\{ e^{-i \int_{\gamma_-(t)} \sum_\alpha \xi_\alpha(\tau_-) A_\alpha(\tau_-)d\tau_-} R(0) \\
    e^{-i \int_{\gamma_+(t)} \sum_\alpha \xi_\alpha(\tau_+) A_\alpha(\tau_+) d\tau_+} \bigg\}.
\end{multline}
From this equation, in which the time arguments are defined on the Keldysh contour, we can go to a physical-time representation
by defining $\xi_\alpha^\pm(\tau) \coloneqq \xi_\alpha(\tau_{\pm})$.
Note that while $\tau_{\pm} \in \gamma_{\pm}(t)$, the variable inside $\xi_\alpha^\pm(\tau)$ is defined between $0$ and $t$.
After changing variable from $\tau_{\pm}$ to $\tau$, we have $d\tau_- = d\tau$, $d \tau_+ = - d \tau$, and
\begin{equation}
    R(t) = G_-(t) R(t_0) G_+^\dagger(t),
\end{equation}
where
\begin{equation} \label{eq:twoside}
    G_\pm(t) \coloneqq \cev{\mathcal{T}} \exp[ -i \sum_\alpha \int_0^t \xi_\alpha^\pm(\tau) A_\alpha(\tau) d\tau ].
\end{equation}
Note that $R(t)$ is not a density operator, since it is not Hermitian.
The rationale of the manipulations done in this section is that we can recover a master equation from Eq. \eqref{eq:twoside} by differentiation:
\begin{equation} \label{eq:twostate}
    \frac{dR(t)}{dt} = -i \sum_\alpha \xi_\alpha^-(t) A_\alpha(t) R(t) + i \sum_\alpha \xi_\alpha^+(t) R(t) A_\alpha(t).
\end{equation}
This is the stochastic master equation we were looking for, and it is the main result of this section.
In case $\varrho(0)$ is a pure state, $\varrho(0) = \dyad{\psi_0}$, we can also write $R(t) = \dyad{\psi_-(t)}{\psi_+(t)}$, where
\begin{equation} \label{eq:twostate_unrav}
    \frac{d\ket{\psi_\pm(t)}}{dt} = -i \sum_\alpha \xi_\alpha^\pm(t) A_\alpha(t) \ket{\psi_\pm(t)}.
\end{equation}
This is sometimes known in the literature as two-state unraveling equation~\cite{stockburger2001diffusion,tilloy2017unraveling}.
We remark that it is possible to obtain Eq. \eqref{eq:twostate_unrav} from the deterministic equation derived in \cite{diosi2014gaussian} (and reobtained in Eq. \eqref{eq:Ferialdi}) by doing a stochastic ansatz, this has been done in \cite{tilloy2017unraveling} and can be thought as an analog in physical time of our Keldysh-based approach leading from \eqref{eq:gaussian_exact} to \eqref{eq:R_solution}.

\begin{figure*}
\includegraphics[width=\textwidth]{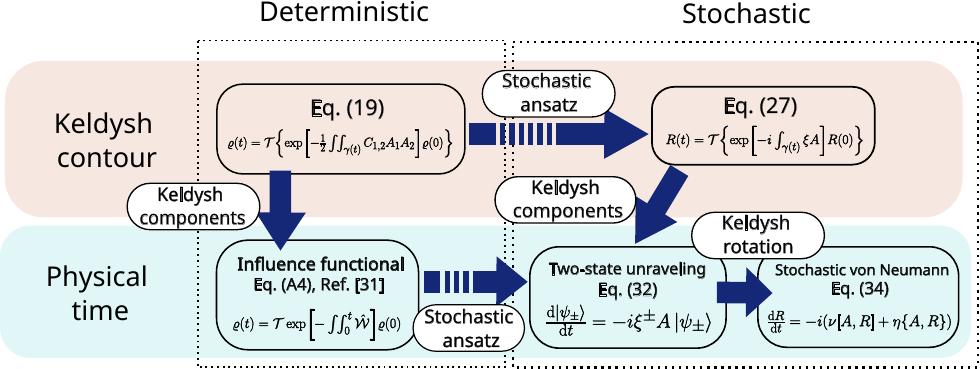}
    \caption{
        A summary of the results obtained from Sec.~\ref{sec:keldysh} to Sec.~\ref{sec:svne}.
        The equations are divided in deterministic and stochastic ones (left and right in the figure, respectively).
        The domain of definition of the time variables of the operator/noises appearing in the equations give an alternative way to classify them.
        This domain can coincide with either the Keldysh contour or the physical time domain (up and down in the figure, respectively).
        Working in the Keldysh formalism, we have shown that the reduced dynamics of a system in contact with a Gaussian bath is described by Eq.~\eqref{eq:gaussian_exact} and in Sec.~\ref{sec:svne} we proved that this can be arranged as the stochastic operator in Eq.~\eqref{eq:R_solution}.
        The connection between the Keldysh contour approach and the physical time approach is always obtained by introducing the Keldysh components (i.e., by splitting $\gamma(t)$ in its forward and backward branches).
        When applied to Eq.~\eqref{eq:gaussian_exact} one recovers the influence functional reported in Ref.~\cite{diosi2014gaussian}; see Eq.~\eqref{eq:Ferialdi} for the explicit expression of the superoperator $\hat{\mathcal W}$.
        The connection between the two-state unraveling~\eqref{eq:twostate_unrav} and the SVNE~\eqref{eq:stocvon} is known~\cite{tanimura2006review,tilloy2017unraveling}: here we showed that it is realized by a Keldysh rotation.
        For completeness, notice that the connection between Eq.~\eqref{eq:Ferialdi} and the two-state unraveling can be done via a Hubbard-Stratonovich technique~\cite{tanimura2006review} or a stochastic ansatz with two noises~\cite{tilloy2017unraveling}.
        Finally, to address cases where the stability condition does not hold, Eq. \eqref{eq:gaussian_exact} and the SVNE must be replaced with Eq. \eqref{eq:gaussian_exact_shifted} and Eq. \eqref{eq:shiftedeq}, respectively.
    }
    \label{fig:scheme}
\end{figure*}

It is also interesting to note that Eq. \eqref{eq:twostate} can be put into a more familiar form by performing the following transformation,
\begin{equation} \label{eq:keldrot}
    \nu_\alpha(t) \coloneqq \frac{\xi_\alpha^-(t) + \xi_\alpha^+(t)}{2},
    \quad
    \eta_\alpha(t) \coloneqq \frac{\xi_\alpha^-(t) - \xi_\alpha^+(t)}{2},
\end{equation}
which is the same as the well-known Keldysh rotation~\cite{kamenev2011book}.
After the transformation, the two-state unraveling reduces to the SVNE~\cite{kubo1969stochastic,tanimura2006review}, which takes the form
\begin{equation} \label{eq:stocvon}
    \frac{dR(t)}{dt} \!=\! -i \sum_\alpha \Big( \nu_\alpha(t) [A_\alpha(t), R(t)] + \eta_\alpha(t) \{ A_\alpha(t), R(t) \} \Big),
\end{equation}
and the newly introduced noises $\nu_\alpha(t)$, $\eta_\alpha(t)$ are easily seen, thanks to Eqs.~\eqref{eq:components}, to be zero-mean Gaussian noises characterized by the following correlation functions:
\begin{gather}
    \mathbb{E}[ \nu_\alpha(\tau_1) \nu_\beta(\tau_2) ] = \Re c_{\alpha\beta}(\tau_1, \tau_2), \notag \\
    \mathbb{E}[ \nu_\alpha(\tau_1) \eta_\beta(\tau_2) ] = i \theta(\tau_1 - \tau_2) \Im c_{\alpha\beta}(\tau_1, \tau_2), \notag \\
    \mathbb{E}[ \eta_\alpha(\tau_1) \eta_\beta(\tau_2) ] = 0.
    \label{eq:rot_corr}
\end{gather}
The first two relations are also known, respectively, as Keldysh and retarded components in the Keldysh formalism literature~\cite{stefanucci2013book,kamenev2011book}.
The advantage of the representation~\eqref{eq:stocvon} with respect to the two-state unraveling~\eqref{eq:twostate_unrav} emerges clearly from the last of Eqs.~\eqref{eq:rot_corr}, that nullifies.
This is a well-known feature of the Keldysh rotation, very commonly used in non-equilibrium field theory \cite{kamenev2011book}.

Note that in order to satisfy the last two equations in~\eqref{eq:rot_corr}, $\eta_\alpha(\tau)$ must be complex.
Instead, the first condition alone is compatible with $\nu_{\alpha}(t)$ being a genuine stochastic process since $ \Re c_{\alpha\beta}(\tau_1, \tau_2) $ is symmetric and positive semidefinite.
This can be seen by writing
\begin{equation} \label{eq:anticomcorr}
    \Re c_{\alpha\beta}(\tau_1,\tau_2) =  \frac{1}{2}
    \Tr[ \{ B_\alpha(\tau_1), B_\beta(\tau_2)\} \Omega ],
\end{equation}
from which the proof follows by noting that for every function $f_{\alpha}(\tau)$ we have
\begin{align} \notag
    \sum_{\alpha,\beta} & \iint_0^t d\tau_1 d\tau_2 \Re c_{\alpha\beta}(\tau_1,\tau_2) f_{\alpha}(\tau_1) f_{\beta}(\tau_2)
    \\
    & = \Tr[ F^2 \Omega ]
    \geq 0
\end{align}
with $F \coloneqq \sum_{\alpha} \int_0^t d\tau f_{\alpha}(\tau) B_{\alpha}(\tau)$.
In addition, note that the imaginary part of the correlation function is independent of the initial state $\Omega$, since
\begin{equation}
    \Im c_{\alpha\beta}(\tau_1,\tau_2) = -   \frac{i}{2}
        \Tr\big[ [B_\alpha(\tau_1), B_\beta(\tau_2)] \Omega \big],
\end{equation}
and the commutator $[B_\alpha(\tau_1), B_\beta(\tau_2)] $ is a $c$-number due to the operators $B_{\alpha}$ being linear in the environment's ladder operators.

Although Eq.~\eqref{eq:stocvon} appeared before in the literature, the interesting point is how our derivation makes it clear that the two noises $\nu_{\alpha}(t)$ and $\eta_{\alpha}(t)$ are actually (the Keldysh-rotated versions of) a single noise that has been split across the Keldysh contour (see also Ref.~\cite{stockburger2002exact}).
It is possible that such insight can yield some advantage during the process of numerically simulating the noise~\cite{lane2020exactly}, in the sense that a single contour noise $\xi_\alpha(z)$ could be easier to generate.
The exploration of this exciting possibility is however out of scope for the present paper, and we leave it for future work.

The stochastic decoupling can also be performed without assuming the stability condition, i.e. starting from Eq.~\eqref{eq:gaussian_exact_shifted}, where now the information about the environment is contained not only in $C_{\alpha,\beta}(z_1,z_2)$ but also in the single-operator average $E_\alpha(z)$.
If we recall Eq.~\eqref{eq:shifted_green}, we can then consider $\xi_\alpha(z)$ to be a Gaussian noise satisfying
\begin{subequations}
    \begin{gather}
        \mathbb{E}[\xi_\alpha(z)] = E_\alpha(z), \label{eq:media1} \\
        \mathbb{E}[ \xi_\alpha(z_1) \xi_\beta(z_2) ] = C_{\alpha,\beta}(z_1,z_2) + E_\alpha(z_1) E_\beta(z_2). \label{eq:varmedia2}
    \end{gather}
\end{subequations}
If we write it in terms of the shifted noise
\begin{equation} \label{eq:shiftmedia}
    \widetilde{\xi}_\alpha(z) \coloneqq \xi_\alpha(z) - E_\alpha(z),
\end{equation}
we obtain the following SVNE:
\begin{multline} \label{eq:shiftedeq}
    \frac{dR(t)}{dt} = -i \sum_\alpha [\nu_\alpha(t) + E_\alpha(t)] [A_\alpha(t), R(t)] \\
    -i \sum_\alpha \eta_\alpha(t) \{A_\alpha(t), R(t)\},
\end{multline}
where
\begin{equation}
    \nu_\alpha(t) = \frac{\widetilde{\xi}^-_\alpha(t) + \widetilde{\xi}^+_\alpha(t)}{2},
    \qquad
    \eta_\alpha(t) = \frac{\widetilde{\xi}^-_\alpha(t) - \widetilde{\xi}^+_\alpha(t)}{2}
\end{equation}
satisfy the same relations we wrote in Eq.~\eqref{eq:rot_corr}.
Note how the shift $E_\alpha(t)$ only influences the commutator part. This implies that it represents a correction to the Hamiltonian component of the dynamics without contributing to dissipation (see also \cite{cavina2024quantum}).
In Fig. \ref{fig:scheme} we do a summary of what we obtained so far, remarking the connections between our method and the existing literature.


\section{Extensions of the formalism}
\label{sec:extensions}

\subsection{Extension to canonically correlated system-bath states}

The assumption~\eqref{eq:uncorrelated} is traditionally employed in the context of the weak-coupling scenario, concurrently with the Born-Markov approximation~\cite{breuer2007book,rivas2010markovian}.
Since we are particularly interested in grasping strong coupling and non-Markovian effects, it may be of interest to instead assume that the system and environment are initially prepared in a (generally correlated) canonical state, that is~\cite{grabert1988quantum,mccaul2017partition,lane2020exactly}
\begin{equation} \label{eq:initial_canonical}
    \varrho_{SE}(0) = \frac{e^{-2b (H'_0 + V)}}{Z},
    \quad
    Z = \Tr[ e^{-2b (H'_0 + V)} ]
\end{equation}
for some $b > 0$ and
\begin{equation}
    H'_0 = H'_S \otimes \mathbbm{1}_E + \mathbbm{1}_S \otimes H_E,
\end{equation}
where $H'_S$ is a ``system preparation'' Hamiltonian, which in general can be different from the bare system Hamiltonian $H_S$ that appears in Eq.~\eqref{eq:H0}. This encompasses the broad framework of quantum quenches, where a quantum system, initially in the  ground state of $H'_S$, begins to evolve after an abrupt change that turns the Hamiltonian into $H_S$ \cite{calabrese2006time,silva2008statistics}).
The value $1/2b$ could be considered as a temperature associated with the initial preparation process (from now on we set $k_B = 1$).

\begin{figure}
    \includegraphics[scale=0.7]{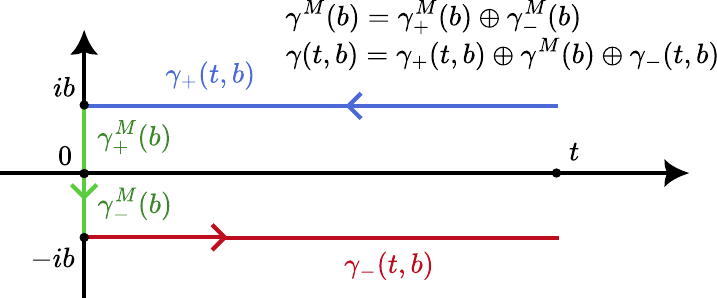}
    \caption{
      The contour $\gamma(t,b)$, obtained by $\gamma(t)$ after shifting $\gamma_\pm(t)$ by $\pm i b$ and after adding a vertical branch $\gamma^M(b)$ running from $ib$ to $-ib$. The key insight for understanding the emergence of the new branch is to interpret the initial correlations between $S$ and $E$ as an effective interaction acting in the imaginary-time (temperature) domain. This interaction, $V$, originates from Eq.~\eqref{eq:initial_canonical} and leads to the interaction-picture operator $\mathcal{V}$ defined in Eqs.~\eqref{eq:exp_corr}, \eqref {eq:interaction_mod} with arguments taken along $\gamma(t,b)$.}
    \label{fig:keldysh2}
\end{figure}

Thanks to the contour formalism, one can show that $\varrho(t)$ can be written in a form that is similar to Eq.~\eqref{eq:ordenv}.
In order to write the result, whose derivation is layed out in Appendix~\ref{app:initial_corr}, we need to consider the following modification of $\gamma(t)$, which arises naturally from the fact that Eq.~\eqref{eq:initial_canonical} can be represented as an imaginary-time evolution.
Specifically, we shift $\gamma_\pm(t)$ by $\pm i b$, calling the result $\gamma_\pm(t,b)$, and we add a vertical branch $\gamma^M(b)$ that runs from $ib$ to $-ib$ (which is similar to the so-called Matsubara branch in the context of many-body physics~\cite{stefanucci2013book,kamenev2011book}).
We call $\gamma^M_+(b)$ the line that goes from $ib$ to 0, and $\gamma^M_-(b)$ the line that goes from 0 to $-ib$.
The whole new contour will be called $\gamma(t,b)$ and is depicted in Fig.~\ref{fig:keldysh2}.
Finally, on this contour we introduce the free propagator
\begin{equation} \label{eq:W0}
    W_0(z_1,z_2) \coloneqq \begin{cases}
        \cev{\mathcal T} \exp[ -i \int_{z_2}^{z_1} H_0(z)dz ], & z_1 \succ z_2, \\
        \vec{\mathcal T} \exp[ i \int_{z_1}^{z_2} H_0(z)dz ], & z_2 \succ z_1,
    \end{cases}
\end{equation}
where $\succ$ is the natural ordering relation on $\gamma(t,b)$, $\cev{\mathcal T}$ and $\vec{\mathcal T}$ are the corresponding contour-ordering and anti-contour-ordering operations, and
\begin{equation} \label{eq:H0_mod}
    H_0(z) \coloneqq \begin{cases}
        H_0 & z \in \gamma_\pm(t,b), \\
        H'_0 & z \in \gamma^M(b).
    \end{cases}
\end{equation}
In Appendix~\ref{app:initial_corr} we then prove that
\begin{equation} \label{eq:exp_corr}
    \varrho_{SE}(t) = \cev{\mathcal T} \qty{ \exp[ -i \int_{\gamma(t,b)} \mathcal{V}(z)dz ] \frac{e^{-2b H'_0}}{Z} },
\end{equation}
where
\begin{equation} \label{eq:interaction_mod}
    \mathcal{V}(z) \coloneqq \begin{cases}
        W_0(-ib,z) V W_0(z,-ib), & \!\!\!z \in \gamma_-(t,b) \oplus \gamma^M_-(b), \\
        W_0(ib,z) V W_0(z,ib), & \!\!\!z \in \gamma^M_+(b) \oplus \gamma_+(t,b).
    \end{cases}
\end{equation}

Eq.~\eqref{eq:exp_corr} is strikingly similar to Eq.~\eqref{eq:keldexp}:
using Eq.~\eqref{eq:interaction_mod}, we can still write a decomposition
\begin{equation}
    \mathcal{V}(z) = \sum_\alpha \mathcal{A}_\alpha(z) \otimes \mathcal{B}_\alpha(z)
\end{equation}
for suitable operators $\mathcal{A}_\alpha(z)$ and $\mathcal{B}_\alpha(z)$, so that
\begin{align} \label{eq:ordenv_mod}
    \varrho(t) &= \frac{1}{Z} \sum_{n=0}^\infty \frac{(-i)^n}{n!} \sum_{\alpha_1,\ldots,\alpha_n} \int_{\gamma(t,b)} dz_1 \ldots dz_n \notag \\
    &\times \Tr[ \cev{\mathcal T}\qty{ \mathcal{B}_1 \ldots \mathcal{B}_n e^{-2b H_E} } ] \cev{\mathcal T} \qty{ \mathcal{A}_1 \ldots \mathcal{A}_n e^{-2b H'_S} },
\end{align}
where again $\mathcal{A}_j \equiv \mathcal{A}_{\alpha_j}(z_j)$, and similarly for $\mathcal{B}_j$.

Eq.~\eqref{eq:ordenv_mod} is analogous to Eq.~\eqref{eq:ordenv}, which, by contrast, was derived under the assumption of uncorrelated initial system-environment state.
Our aim is to generalize the results of Secs. \ref{sec:gaussian} and \ref{sec:svne} to the present scenario, i.e. we look for
an analog of Eqs. \eqref{eq:gaussian_exact}, \eqref{eq:gaussian_exact_shifted}, \eqref{eq:R_solution} that hold in this case.
The generalization of these results becomes straightforward upon recognizing two key observations:
\begin{enumerate}
    \item The assumption of a Gaussian environment allows the application of Wick's theorem, regardless of the contour on which the variables \( z_1, \ldots, z_n \) are defined \cite{stefanucci2013book};
    \item The stochastic decoupling introduced in Sec.~\ref{sec:svne} can be readily extended, provided that the new stochastic variable \( \xi_{\alpha}(z) \) is defined on the updated contour.
\end{enumerate}
Given this, we can apply the strategy of secs. \eqref{sec:gaussian} and \eqref{sec:svne} to Eq. \eqref{eq:ordenv_mod} and obtain that the evolved system density matrix can be recovered by averaging the operator
\begin{equation} \label{eq:R_solutionCorr}
    R_b(t) = \cev{\mathcal{T}} \qty{ \exp[ -i \sum_\alpha \int_{\gamma(t,b)} \xi_\alpha(z) \mathcal A_\alpha(z) dz ] R(0) },
\end{equation}
where the noises $\xi_{\alpha}(z)$ now take value on $\gamma(t,b)$ and satisfy the condition \eqref{eq:corranz} on such contour.

Note that a generalization of the stochastic von Neumann equation can be derived from~\eqref{eq:R_solutionCorr} (see the discussion at the end of Sec.~\ref{sec:keldysh}).
Seminal works have already accomplished this using path-integral techniques~\cite{grabert1988quantum,mccaul2017partition,lane2020exactly}.

\subsection{Application to quantum thermodynamics}

The application to correlated initial states demonstrates the versatility of our method. Specifically, given an equation of the form \eqref{eq:vNeu_sol_keldysh}, valid on any contour, we can derive compact expressions such as \eqref{eq:gaussian_exact}, \eqref{eq:R_solution} for suitable noises $\xi_{\alpha}(z)$ defined on the contour of interest.

\begin{figure}
    \includegraphics[scale=0.7]{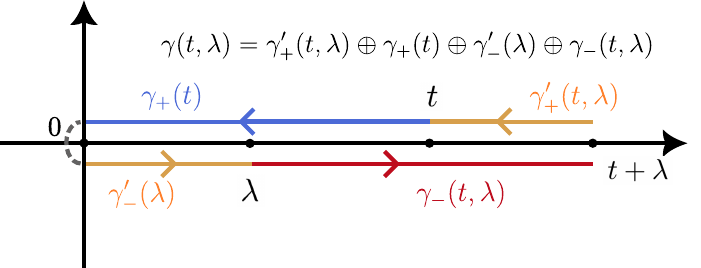}
    \caption{   The contour $\gamma(t,\lambda)$, obtained by $\gamma(t)$ after adding two additional horizontal branches of length $\lambda$ before $\gamma_+$ and $\gamma_-$. This contour is a powerful tool to study the statistics in TPEM schemes \cite{funo2018pathint}.}
    \label{fig:keldysh3}
\end{figure}

In this section, we apply this principle to derive a stochastic equation that describes energy statistics in quantum thermodynamics. We adopt the standard assumption that work corresponds to the stochastic outcome of a two-point energy measurement (TPEM) protocol: the system undergoes unitary evolution governed by the total Hamiltonian given in \eqref{eq:totH}, and measurements of $H(\tau)$ are performed both before and after this evolution \cite{Espositoreview}.
Similarly, the heat can be obtained by performing a double measurement of $H_E$ instead of $H(\tau)$.
Under this assumption, the characteristic function of the work (heat) statistics—i.e., the statistics of the difference between the two measurement outcomes—takes the form:
\begin{equation} \label{eq:MGF}
    M_{\Lambda}(t,\lambda) = \Tr \big[e^{i \lambda \Lambda(t)} U(t,0) \rho_{SE}(0)  e^{- i \lambda \Lambda(0)} U^{\dag}(t,0)  \big],
\end{equation}
where $\Lambda(t) = H(t), \Lambda(0) =H(0)$ for a work TPEM and $\Lambda(0), \Lambda(t) = H_E$ for a heat TPEM, while  $\lambda$ is a counting parameter that allows us, under differentiation, to recover the different moments of the probability distributions.
It is well known that the expression above can be formulated in terms of an ordered exponential on a modified Keldysh contour \cite{cavina2023convenient, funo2018pathint}.
Intuitively, we can treat the two complex exponentials in $\lambda$ as ``virtual" time evolutions by augmenting the forward and backward branches with two additional tracks of length $\lambda$, that we call $\gamma_+'(\lambda)$ and $\gamma_-'(t,\lambda)$.
Accordingly, the measurement operators $\Lambda(0), \Lambda(t)$ are treated as the values of an extended Hamiltonian in the tracks $\gamma_-'(t,\lambda)$ and $\gamma_+'(\lambda)$, respectively.
This originates the contour $\gamma(t,\lambda)$ in Fig. \ref{fig:keldysh3}, while the extended Hamiltonian $H(z)$ reads

\begin{equation} \label{eq:H0_mod2}
    H(z) \coloneqq \begin{cases}
        \Lambda(t) & z \in \gamma_+'(t,\lambda), \\
        H(\tau) & z = \tau \in \gamma_+(t),
        \\
        \Lambda(0) & z \in \gamma_-'(\lambda), \\
        H(\tau-\lambda) & z = \tau \in \gamma_-(t, \lambda),
    \end{cases}
\end{equation}

where we also admitted the possibility of a time-dependent physical Hamiltonian $H(\tau)$, for the sake of generality.
We can reformulate the moment generating function as the trace of an ordered exponential on the contour $\gamma(t,\lambda)$, that is
\begin{equation} \label{eq:countord1}
    M_{\Lambda}(t,\lambda) = \Tr \cev{\mathcal{T}}  \qty{  \exp [
 \int_{\gamma(t,\lambda)} H(z) dz ]  \rho_{SE}(0) } ,
\end{equation}
where $H(z)$ is the one given in Eq.~\eqref{eq:H0_mod2}.
This generalizes Eq.~\eqref{eq:vNeu_sol_keldysh} to the case considered in this section.
Assuming the Gaussian environment hypothesis, we can eliminate the degrees of freedom of the environment and compute $\rho(t,\lambda) \coloneqq Tr_E[\rho_{SE}(t,\lambda)]$, where $\rho_{SE}(t,\lambda)$ is the ordered exponential in Eq. \eqref{eq:countord1}, and follow the steps of Sec. \ref{sec:gaussian}.
We do it explicitly for the heat generating function, i.e. we assume
\begin{align} \notag
     H_0(t) & = H_S(t) \otimes \mathbbm{1}_E + \mathbbm{1}_S \otimes H_E, \label{eq:H0new} \\
        V & = \sum_\alpha A_\alpha \otimes B_\alpha, \notag \\
        \Lambda(0)& = \Lambda(t) = H_E.
\end{align}

The calculations are exactly the same as the ones carried out in Sec. \ref{sec:gaussian} while a stochastic decoupling of the bath can be obtained by following the same procedures of Sec. \ref{sec:svne}. We can just copy the results of those sections, and conclude that $\rho(t,\lambda)$ can be obtained by averaging the stochastic operator

\begin{equation} \label{eq:R_solutionlamb}
    R(t,\lambda) = \cev{\mathcal{T}} \qty{ \exp[ -i \sum_\alpha \int_{\gamma(t)} \xi_{\lambda,\alpha}(z) \mathcal{A}_{\alpha}(z) dz ] R(0) },
\end{equation}

where $\mathcal{A}_{\alpha}(z) = A_{\alpha}(\tau)$ for $z = \tau_{\pm}$ and the noise $\xi_{\alpha, \lambda}(z)$ bears an additional dependence by $\lambda$ since it has to satisfy the condition \eqref{eq:corranz} for all $z_1,z_2 \in \gamma(t,\lambda)$. Notice that the integration in Eq.~\eqref{eq:R_solutionlamb} is only on $\gamma(t)$ since the interaction Hamiltonian $V$ is non-zero only on this contour.
We can perform the Keldysh rotation and obtain a stochastic equation in the form of a SVNE for $R(t,\lambda)$.
Using again the formal analogy between Eq.~\eqref{eq:R_solutionlamb} and Eq.~\eqref{eq:R_solution} is immediate to derive

\begin{align} \notag
    \frac{dR(t,\lambda)}{dt} & \!=\! -i \sum_\alpha \Big( \nu_{\lambda, \alpha}(t) [A_\alpha(t), R(t)]  \\ & +  \eta_{\lambda, \alpha}(t) \{ A_\alpha(t), R(t) \} \Big), \label{eq:stocvonlambda}
\end{align}
where $\nu_{\lambda, \alpha}(t), \eta_{\lambda, \alpha}(t)$ are obtained by extracting the forward/backward components of $\xi_{\lambda, \alpha}(z)$ and performing a Keldysh rotation.
The extended correlation functions of these two noises are explicitly computed in App. \ref{sec:therm}. We remark that for $\lambda \neq 0$ the autocorrelation of $\eta_{\lambda, \alpha}(t)$ does not nullify, differently from what happens in the standard case treated in Sec. \ref{sec:svne}.
The equation above opens the way to a possible new approach for numerically computing TPEM characteristic functions by stochastically simulating trajectories on the modified contour shown in Fig. \ref{fig:keldysh3}. A detailed analysis of this approach is reserved for future work.

\section{\label{sec:reduction}
    Reduction to a single physical-time noise
}

In Sec.~\ref{sec:svne} we saw how the dynamics can be described by a stochastic master equation that is written in terms of a Keldysh noise $\xi_\alpha(z)$ or, equivalently, in terms of two physical-time noises $\nu_\alpha(\tau)$ and $\eta_\alpha(\tau)$.
A possible advantage of stochastic unraveling techniques is that one can approach numerical problems by generating trajectories for the noises.
These simulations are typically efficient, but some issues can arise due to the fact that the dynamics generated by the SVNE does not preserve Hermiticity~\cite{stockburger2004simulating,zhou2005stochastic}.
The presence of the noises $\eta_{\alpha}(t)$ in Eq.~\eqref{eq:stocvon} is at the heart of this problem, as discussed at the end of Sec.~\ref{sec:svne}.
In this section we show that, using the specific form of Eq.~\eqref{eq:rot_corr}, we can actually describe the dynamics in terms of $\nu_\alpha(\tau)$ alone.
However, this will come at a cost, since after the elimination of $\eta_{\alpha}(\tau)$ it will not be immediately obvious how to express the result as the solution of a time-local differential equation.

Let us start from Eq.~\eqref{eq:R_solution}, which is the solution of the stochastic von Neumann equation in~\eqref{eq:stocvon}.
If we perform the Keldysh rotation \eqref{eq:keldrot}, in the form $\xi_\alpha^-(\tau) = \nu_\alpha(\tau) + \eta_\alpha(\tau)$, $\xi_\alpha^+(\tau) = \nu_\alpha(\tau) - \eta_\alpha(\tau)$, we can write
\begin{multline} \label{eq:RAC}
    R(t) = \cev{\mathcal{T}} \bigg\{ \exp \bigg[ -i\sum_\alpha \int_0^t d\tau \\
    \big( \nu_\alpha(\tau) \mathcal{C}_\alpha(\tau) + \eta_\alpha(\tau) \mathcal{A}_\alpha(\tau) \big) d\tau \bigg] R(0) \bigg\},
\end{multline}
where we introduced the ``commutator'' and ``anticommutator'' operators as
\begin{subequations}
    \begin{gather}
        \mathcal{C}_\alpha(\tau) \coloneqq A_\alpha(\tau_-) - A_\alpha(\tau_+), \\
        \mathcal{A}_\alpha(\tau) \coloneqq A_\alpha(\tau_-) + A_\alpha(\tau_+).
    \end{gather}
\end{subequations}
In order to expand the exponential in Eq.~\eqref{eq:RAC}, we must be able to write products of sums of operators as sums of products.
To do that, we use the following fact: given two generic collections of operators $\{X_i\}_{i=1}^n$, $\{Y_i\}_{i=1}^n$ we can write
\begin{multline} \label{eq:operator_binomial}
    \prod_{i=1}^n \qty(X_i + Y_i) = \sum_{k=0}^n \frac{1}{k!(n-k)!} \\
    \times \sum_{\sigma \in \mathfrak{S}_n} \mathcal{I} \qty{ X_{\sigma(1)} \ldots X_{\sigma(k)} Y_{\sigma(k+1)} \ldots Y_{\sigma(n)} },
\end{multline}
where $\mathfrak{S}_n$ is the set of permutations of $\{1,\ldots,n\}$ and $\mathcal{I}$ is an index-ordering operation, which puts operators with lower index on the left.
In our case, $X_i \mapsto \nu_\alpha(\tau_i) \mathcal{C}_\alpha(\tau_i)$ and $Y_i \mapsto \eta_\alpha(\tau_i) \mathcal{A}_\alpha(\tau_i)$, where $\tau_i$ is one of the time variables that appear when expanding Eq.~\eqref{eq:RAC}.
Notice also how the index ordering $\mathcal{I}$ naturally turns into a time ordering $\cev{\mathcal{T}}$ in this context.
Moreover, thanks to the integration structure and the fact that operators commute inside the $\cev{\mathcal{T}}$ sign, every permutation in Eq.~\eqref{eq:operator_binomial} contributes equally, so that $\sum_{\sigma \in \mathfrak{S}_n}$ can be dropped if we multiply by $n!$.
As a consequence,
\begin{multline} \label{eq:gigante}
    R(t) = \sum_{n=0}^\infty \frac{(-i)^n}{n!} \sum_{\alpha_1,\ldots,\alpha_n} \int_0^t d\tau_1 \ldots d\tau_n \sum_{k=0}^n \binom{n}{k} \\
    \nu_1 \ldots \nu_k \eta_{k+1} \ldots \eta_n \cev{\mathcal{T}} \qty{ \mathcal{C}_1 \ldots \mathcal{C}_k \mathcal{A}_{k+1} \ldots \mathcal{A}_n R(0) },
\end{multline}
where, to simplify notations, we defined $\nu_j \equiv \nu_{\alpha_j}(\tau_j)$ and similarly for $\eta_j$, $\mathcal{C}_j$, and $\mathcal{A}_j$.

Now we remember that $R(t)$ can be arbitrarily modified as long as we preserve the property $\mathbb{E}[R(t)] = \varrho(t)$.
Let then us consider the expectation of Eq.~\eqref{eq:gigante}, which only affects the string of noises.
Since $\nu_\alpha(\tau)$ and $\eta_\alpha(\tau)$ are correlated Gaussian noises, we can use the scalar Wick's theorem as we did in Eq.~\eqref{eq:isserlis}.
However, the conditions~\eqref{eq:rot_corr} greatly simplify the pairings structure.
In particular, we must require that every occurrence of $\eta_\alpha(\tau)$ matches a corresponding occurrence of $\nu_\alpha(\tau)$; the remaining occurrences of $\nu_\alpha(\tau)$ are then free of matching with each other.
For this to happen it is necessary to have $n$ even, say $n = 2m$, and $k \geq m$.
In Appendix~\ref{app:partial_wick} we show that for the sake of calculating Eq.~\eqref{eq:gigante} one can make the substitution
\begin{multline} \label{eq:partial_isserlis}
    \mathbb{E}[\nu_1 \ldots \nu_k \eta_{k+1} \ldots \eta_{2m}] \mapsto \theta(k-m) i^{2m-k} \frac{k!}{(2k-2m)!} \\
    G_{1,k+1} \ldots G_{2m-k,2m} \mathbb{E}[\nu_{2m-k+1} \ldots \nu_k],
\end{multline}
where
\begin{equation}
    G_{i,j} \coloneqq -i \mathbb{E}[\nu_i \eta_j] = \theta(\tau_i-\tau_j) \Im c_{\alpha_i,\alpha_j}(\tau_i,\tau_j).
\end{equation}
We can now define a new stochastic operator $R'(t)$ that is obtained after removing the remaining expectation value sign around the product of $\nu$ noises:
\begin{multline} \label{eq:Rtilde}
    R'(t) = \sum_{m=0}^\infty \sum_{k=m}^{2m} \frac{(-i)^k}{k!} \binom{k}{2m-k} \sum_{\alpha_1,\ldots,\alpha_{2m}} \\
    \int_0^t d\tau_1 \ldots d\tau_{2m} G_{1,k+1} \ldots G_{2m-k,2m} \nu_{2m-k+1} \ldots \nu_k \\
    \cev{\mathcal{T}} \qty{ \mathcal{C}_1 \ldots \mathcal{C}_k \mathcal{A}_{k+1} \ldots \mathcal{A}_{2m} R'(0) },
\end{multline}
where $R'(0) \equiv R(0)$.
By construction, we still have $\mathbb{E}[R'(t)] = \mathbb{E}[R(t)]$, and the dynamics is equivalently and completely described by using $R'(t)$ instead of $R(t)$.
The important point here is that Eq.~\eqref{eq:Rtilde} does not explicitly depend on the noise $\eta_\alpha(\tau)$ and the noise $\nu_{\alpha}(\tau)$ has positive semidefinite autocorrelation.

An important question is whether it is possible to find a master equation that describes the evolution of $R'(t)$.
Such an equation would be an exact non-Markovian stochastic equation which is written in terms of a single real Gaussian noise, in contrast to more common diffusion equations, which are instead described by a complex Gaussian noise~\cite{diosi1998diffusion,stockburger2001diffusion}.
We leave this remarkable possibility to a future work.


\section{\label{sec:measurements}
    Quantum noises and measurements of the environment
}

The noises $\nu_{\alpha}(t)$ and $\eta_{\alpha}(t)$ appearing in the stochastic von Neumann equation~\eqref{eq:stocvon} are generally complex valued, even though we saw in Sec.~\ref{sec:reduction} that there is a representation of the dynamics in which $\eta_\alpha(t)$ can be ``eliminated'' and $\nu_\alpha(\tau)$ can be real valued.
Unfortunately, (the real part of) $\nu_{\alpha}(t)$ cannot be interpreted \textit{a priori} as the outcome of a continuous measurement process performed on the environment~\cite{gambetta2002measurement,diosi2008retarded,wiseman2008pure,megier2020measurement, kronke2012non}.
This contrasts with what occurs in other stochastic equations used in quantum metrology~\cite{albarelli2024pedagogical} and quantum thermodynamics~\cite{manzano2018quantum}, which are written in terms of the stochastic measurement records.
This is the case, for instance, in homodyne, heterodyne, and photodetection master equations, in which the noise is associated with the outcomes of weak measurements of the position, momentum, and photon number operators.

In this last section we will analyze how a one-shot measurement of the environment performed at the beginning of the dynamics affects the noises $\nu_{\alpha}(t)$ and $\eta_{\alpha}(t)$.
In addition, we will focus on finding a measurement protocol and a physical regime for the environment in which the measurement records $\mathbf{y}$ will give us some scalar stochastic estimate $b_{\alpha}^{(\mathbf y)}(t)$ of the operator $B_{\alpha}(t)$ with the same autocorrelation function as the noise $\nu_{\alpha}(t)$, meaning that
\begin{equation} \label{eq:cond1}
    \mathbb{E}_{\nu}[ \nu_{\alpha}(\tau_1) \nu_{\beta}(\tau_2) ] = \mathbb{E}_{y}[ b_{\alpha}^{(\mathbf y)}(\tau_1) b_{\beta}^{(\mathbf y)}(\tau_2) ].
\end{equation}
Here, the symbol $\mathbb{E}_\nu$ on the left-hand side stands for an average with respect to the noises $\nu_\alpha$, while the symbol $\mathbb{E}_y$ stands for an average over the measurement outcomes $\mathbf{y}$.
Additionally, to ensure that the dynamics after the measurement remains consistent with the one without it, we will require that averaging over $\mathbf{y}$ will eventually restore the original density matrix.
In other words, the system dynamics should not be altered by the collapse induced by the initial environment measurement.
These conditions grant that, in some cases, at least the real part of $\nu_\alpha(\tau)$ can be interpreted as a ``classical noise'' emerging from a measurement scheme.


\subsection{\label{sec:effmeas}
    Effect of an initial measurement on the stochastic von Neumann equation
}

Suppose the environment is in a state $\Omega$ and a generic POVM~\cite{watrous2018book} with measurement operators $\{M_y\}$ is performed, so that $y$ is the outcome of the measurement, $\sum_y M_y^\dagger M_y = \mathbbm{1}$, and the post-measured state $\Omega_y$, conditioned on obtaining the outcome $y$, writes
\begin{equation} \label{eq:after}
    \Omega_y = \frac{M_y \Omega M^\dagger_{y}}{\mathbb{P}(y)},
    \quad
    \mathbb{P}(y) = \Tr[M_y \Omega M^\dagger_y],
\end{equation}
where $\mathbb{P}(y)$ is the probability associated with the outcome $y$.
Combining several POVMs results in a measurement record $\mathbf{y} = (y_1, \ldots, y_n)$ and in an associated post-measured state
\begin{equation} \label{eq:after_multiple}
    \Omega_{\mathbf y} = \frac{M_{\mathbf y} \Omega M^\dagger_{\mathbf y}}{\mathbb{P}(\mathbf y)},
    \quad
    \mathbb{P}(\mathbf y) = \Tr[M_{\mathbf y} \Omega M^\dagger_{\mathbf y}],
\end{equation}
where $M_{\mathbf y} = M_{y_n} \ldots M_{y_1}$.

Let us now imagine the following scenario: given an environment in the initial Gaussian state $\Omega$, we perform an arbitrary finite number of measurements on it before coupling it to the system.
For a measurement record $\mathbf{y}$, the state $\Omega$ is then replaced by $\Omega_{\mathbf y}$ in Eq.~\eqref{eq:after_multiple}.
In case $\Omega_{\mathbf y}$ is still a Gaussian state, when coupling the system to the post-measured environment we can apply all the machinery developed in Secs.~\ref{sec:gaussian}, \ref{sec:svne} and \ref{sec:reduction}, with the only caveat that the correlation functions in Eqs.~\eqref{eq:cGF}, \eqref{eq:physcorr} will change.
The latter (from which all the GFs used in this manuscript can be derived, see Eqs.~\eqref{eq:components}, \eqref{eq:rot_corr}) is indeed replaced by
\begin{equation}  \label{eq:newcorr}
    c_{\alpha \beta}^{(\mathbf y)} (\tau_1,\tau_2) \coloneqq \Tr[  B_{\alpha}(\tau_1) B_{\beta}(\tau_2) \Omega_{\mathbf y} ] -  E^{(\mathbf y)}_{\alpha}(\tau_1)  E^{(\mathbf y)}_{\beta}(\tau_2),
\end{equation}
where, as in Eq.~\eqref{eq:shifted_green}, we accounted for a possible shift due to the non-vanishing average of the post-measured state
\begin{equation} \label{eq:newavg}
   E^{(\mathbf y)}_{\alpha}(\tau) \coloneqq \Tr[B_{\alpha}(\tau) \Omega_{\mathbf y} ] \neq 0.
\end{equation}
Indeed, even if $\Omega$ satisfies the stability condition~\eqref{eq:stability}, in general this is not the case for the post-measured state $\Omega_{\mathbf y}$.
We can thus rely on Eq.~\eqref{eq:shiftedeq} to write the stochastic von Neumann equation
\begin{align} \notag
    \frac{dR_{\mathbf y}(t)}{dt} & = -i \sum_\alpha \big[ \nu^{(\mathbf y)}_\alpha(t) + E^{(\mathbf y)}_{\alpha}(t) \big] [A_\alpha(t), R_{\mathbf y}(t)] \\ & - i \sum_\alpha \eta^{(\mathbf y)}_\alpha(t) \{ A_\alpha(t), R_{\mathbf y}(t) \},  \label{eq:stocvon2}
\end{align}
where $R_{\mathbf y}(t)$ is the system stochastic operator associated with this modified evolution and where $\nu^{(\mathbf y)}_\alpha(t)$, $\eta^{(\mathbf y)}_\alpha(t)$ are zero-average Gaussian noises such that
\begin{gather}
    \mathbb{E}_\nu[ \nu^{(\mathbf y)}_\alpha(\tau_1) \nu^{(\mathbf y)}_\beta(\tau_2) ] = \Re c^{(\mathbf y)}_{\alpha\beta}(\tau_1, \tau_2), \notag \\
    \mathbb{E}_\nu[ \nu^{(\mathbf y)}_\alpha(\tau_1) \eta^{(\mathbf y)}_\beta(\tau_2) ] = i \theta(\tau_1 - \tau_2) \Im c_{\alpha\beta}(\tau_1, \tau_2), \notag \\
    \mathbb{E}_\nu[ \eta^{(\mathbf y)}_\alpha(\tau_1) \eta^{(\mathbf y)}_\beta(\tau_2) ] = 0. \label{eq:rotcorr2}
\end{gather}
Note that in Eq.~\eqref{eq:stocvon2} there are two sources of stochasticity at play: first, the fluctuation of the measurement record $\mathbf{y}$, and second, the fluctuation of the noises $\nu_\alpha^{(\mathbf y)}(t)$ and $\eta_\alpha^{(\mathbf y)}(t)$ at fixed $\mathbf{y}$, derived from the stochastic decoupling.
The averages appearing in Eq.~\eqref{eq:rotcorr2} are intended with respect to this second source only.
The stochasticity of $\mathbf{y}$ influences instead both the second moment of $\nu_\alpha^{(\mathbf y)}(t)$ and the bias $E_\alpha^{(\mathbf y)}(t)$.
Note that the second and the third correlation functions in Eq.~\eqref{eq:rotcorr2} remain untouched by the measurement [and are the same as in Eq.~\eqref{eq:rot_corr}], since they are independent from the initial state of the environment (see the discussion at the end of Sec. \ref{sec:svne}).


\subsection{\label{sec:suffmeas}
    Sufficient conditions for a measurement interpretation
}

To find the conditions under which a generic quantum measurement scheme yields Eq.~\eqref{eq:cond1}, we will narrow our field of analysis by making some assumptions.
First, we take $B_{\alpha} \equiv X_{\alpha}$, where $\alpha$ can be thought of as a label running over the modes of the environment, and $X_{\alpha}$ is the position operator associated with mode $\alpha$.
We will also assume that each mode $\alpha$ is a harmonic oscillator of frequency $\omega_{\alpha}$ and mass $m_{\alpha}$.

In addition, two hypotheses on the measurement protocol will be made.
To preserve the form of the Gaussian stochastic equations, we need Gaussian post-measured states (see Sec. \ref{sec:effmeas}); therefore, we will consider only measurements described by the following measurement operators:
\begin{equation} \label{eq:M_Y}
    M_{Y,y} = \frac{1}{\qty(2\pi \sigma_Y^2)^{1/4}} \exp[ -\frac{(y-Y)^2}{4 \sigma_Y^2} ].
\end{equation}
Specifically, $M_{Y,y}$ is the measurement operator associated with a measurement of the observable $Y$ with obtained outcome $y$.
The quantity $\sigma_Y^{-1}$ can be interpreted as the precision with which such a measurement is carried out.

Secondly, we adopt a heterodyne scheme in which position and momentum of each environmental mode is measured at the initial time, resulting in [see Eq.~\eqref{eq:after_multiple}]
\begin{equation} \label{eq:omega_y}
    \Omega_{\mathbf{x}.\mathbf{p}} = \frac{M_{\mathbf{X},\mathbf{x}} M_{\mathbf{P},\mathbf{p}} \Omega M^\dagger_{\mathbf{P},\mathbf{p}} M^\dagger_{\mathbf{X},\mathbf{x}}}{\mathbb{P}(\mathbf{x},\mathbf{p})},
\end{equation}
where $M_{\mathbf{X},\mathbf{x}} = \prod_\alpha M_{X_\alpha,x_\alpha}$ (the ordering of the factors is irrelevant for our purposes), $M_{\mathbf{P},\mathbf{p}} = \prod_\alpha M_{P_\alpha,p_\alpha}$, with $P_\alpha$ being the momentum operator associated with mode $\alpha$, and $\mathbb{P}(\mathbf{x},\mathbf{p})$ is the probability of obtaining the strings of outcomes $\mathbf{x}$ for the position and $\mathbf{p}$ for the momentum.
For simplicity, we will indicate $\mathbf{y} = (\mathbf{x}, \mathbf{p})$ and $y_\alpha = (x_\alpha, p_\alpha)$.
After these initial measurements, the environment is coupled to the system, and the subsequent system evolution is governed by Eqs.~\eqref{eq:stocvon2} and~\eqref{eq:rotcorr2}.

Finally, let us assume without much loss of generality that $\Tr[X_\alpha \Omega] = \Tr[P_\alpha \Omega] = 0$.
Since $\Omega$ is a Gaussian state, it is then completely characterized by the variances
\begin{equation} \label{eq:delta_variances}
    \Delta_{X_\alpha}^2 \coloneqq \Tr[X_\alpha^2 \Omega],
    \quad
    \Delta_{P_\alpha}^2 \coloneqq \Tr[P_\alpha^2 \Omega].
\end{equation}

Using the Wigner function formalism~\cite{polkovnikov2010phase}, in Appendix~\ref{app:wigner} we show that if the following conditions hold,
\begin{subequations} \label{eq:wigner_conditions}
    \begin{gather}
        \sigma_{X_\alpha} \ll \Delta_{X_\alpha},
        \quad
        \sigma_{P_\alpha} \ll \Delta_{P_\alpha},
        \label{eq:sigma_less_delta} \\
        \sigma_{X_\alpha} \sigma_{P_\alpha} \gg \hbar,
        \label{eq:sigma_uncertainty}
    \end{gather}
\end{subequations}
then
\begin{subequations} \label{eq:wigner_result}
    \begin{gather} \label{eq:wigner_result1}
        E_\alpha^{(\mathbf{y})}(t) \simeq x_\alpha \cos(\omega_\alpha t) + \frac{p_\alpha}{m_\alpha \omega_\alpha} \sin(\omega_\alpha t), \\
        \mathbb{E}_y \qty[ E_\alpha^{(\mathbf{y})}(\tau_1)  E_\beta^{(\mathbf{y})}(\tau_2) ] \simeq \mathbb{E}_\nu \qty[ \nu_\alpha(\tau_1) \nu_\beta(\tau_2) ], \label{eq:wigner_result2}
    \end{gather}
\end{subequations}
where in Eq.~\eqref{eq:sigma_uncertainty} we restored the dependence on $\hbar$ for dimensional clarity and by comparison with \eqref{eq:cond1} we have $b_{\alpha}^{(\mathbf{y})} \coloneqq E_{\alpha}^{(\mathbf{y})}$. Notice that $\nu_\alpha(\tau)$ is the noise we had without performing the initial measurements.
We now briefly discuss the physical interpretation of the conditions found above.
Note that combining  Eqs.~\eqref{eq:wigner_conditions} together implies

\begin{equation} \label{eq:inspread}
    \Delta_{X_\alpha} \Delta_{P_\alpha} \gg \hbar.
\end{equation}

This means that the fluctuations of the initial preparation $\Omega$ are characterized by position and momentum scales that are so large that we can neglect any quantum effect. Under these assumptions, the initial Wigner function associated to $\Omega$ can be thought of as a classical phase-space probability distribution.
By choosing the precision properly as in Eq. \eqref{eq:sigma_less_delta}
the distribution of the outcomes coincides with the probability distribution representing the initial state at time $0$.
Then, Eq. \eqref{eq:wigner_result1} follows from  $E_{\alpha}^{(\mathbf{y})}(0) = x_{\alpha}$
and Eq. \eqref{eq:wigner_result2} arises from the phase-space statistics of the initial state matching the statistics of the measurement outcomes (a detailed derivation of this fact is contained in App. \ref{app:wigner}).

We conclude by mentioning what happens to Eq.~\eqref{eq:stocvon2} in the regime identified by the conditions \eqref{eq:wigner_conditions}.
The measurement is very precise, so that the post-measured state
$\Omega_{\mathbf y}$ has very small position and momentum variances (respectively given, for any assigned mode $\alpha$, by $\sigma_{X_{\alpha}}^2, \sigma_{P_{\alpha}}^2$).
These are much smaller than the variances associated with the original
state preparation, that, following Eq. \eqref{eq:wigner_result2}, are now determining the fluctuations of
$E_\alpha^{(\mathbf{y})}(t)$.
Thus, the fluctuations of $\nu_{\alpha}^{(\mathbf{y})}$ in Eq. \eqref{eq:stocvon2} are very small if compared to the ones of $E_\alpha^{(\mathbf{y})}(t)$ and the fluctuations of the Hamiltonian part are totally controlled by the latter \footnote{Note that $\nu_{\alpha}^{\mathbf{y}}$ gives no sensible contributions to the fluctuations, still we cannot eliminate it from the equations since it contributes through its coupling with $\eta_{\beta}^{(\mathbf{y})}$ in the second of Eqs. \eqref{eq:rotcorr2}}.
Hence, we identified a regime where the first correlation function in Eq. \eqref{eq:rot_corr} can be interpreted as stemming from the stochasticity of  outcomes of an appropriately chosen measurement of the environment.


\section{\label{sec:conclusions}
    Conclusions
}

We gave a unified derivation of various stochastic quantum equations for the exact non-Markovian evolution of open quantum systems by using a new technique based on the application of the Wick's theorem on ordered contours.
In one of our fundamental equations [Eq.~\eqref{eq:twostate}] the effect of the environment is encoded through a single stochastic field $\xi_\alpha(z)$ defined on the Keldysh contour; when decomposed in its physical-time components with a Keldysh rotation, the well-known stochastic von Neumann equation (SVNE) is obtained [see Eq.~\eqref{eq:stocvon}].
Although Eq. \eqref{eq:stocvon} has already been derived and extensively studied in the literature, our approach constitutes an alternative to other common techniques that have been employed before (such as path integration) and allows for an easy generalization of the SVNE to three important cases: (i) when the stability condition does not hold [see Eq. \eqref{eq:gaussian_exact_shifted}], (ii) when there are initial correlations between the system and the environment [see Eq. \eqref{eq:R_solutionCorr}], and (iii) when calculating the generating function in a TPEM scenario [see Eq. \eqref{eq:stocvonlambda}].
Thanks to the perspective offered by our technique, we noted that the symmetric component $\nu_\alpha(t)$ of the Keldysh-rotated noise is all that one needs to characterize the solution of the SVNE, even though more work is needed to find a corresponding stochastic master equation. Moreover, at least in the semiclassical limit of an environment in a state with very broad position and momentum uncertainty, we managed to give a partial operational meaning to such noise.

Our results pave the way for several interesting developments.
The Keldysh approach can be used in similar contexts in which we have a different statistics (for instance, in the case of a system coupled to fermionic baths~\cite{cirio2022influence}) and it is also a natural choice if we are interested in studying thermodynamic-related symmetries, like the the Fluctuation Theorem~\cite{aron2018non, sieberer2015thermodynamic,cavina2023convenient}.
Given the symmetric structure of Eq.~\eqref{eq:gaussian_exact}, the Keldysh representation could also be a natural framework in which to study regularizations of nonpositive master equations, such as the Redfield equation~\cite{nathan2020ule,mozgunov2020positive,mcCauley2020lindblad,dabbruzzo2023reg,dabbruzzo2024choi}. Studies in this direction were already performed in Ref.~\cite{reimer2019density}, but it could be possible to specialize them to the Gaussian environment scenario considered here.
In the broader context of quantum master equations, the techniques developed in this work can also serve as a foundation for deriving a general non-Markovian master equation for a Gaussian system coupled to a Gaussian environment \cite{d2025exact}.
Another promising direction is the generalization of our calculation to other contours:
as already observed in other contexts~\cite{stefanucci2013book,ferialdi2021wick}, the validity of the Wick's theorem is not related to the shape of the contour.
We already had a taste of this phenomenon when we discussed the generalization of the SVNE to the case of an environment initially correlated with the system (going from the contour in Fig. \ref{fig:keldysh} to the contour in Fig. \ref{fig:keldysh2}) and to the case of  heat statistics (the contour in Fig. \ref{fig:keldysh3}, see also \cite{cavina2024quantum, cavina2023convenient, funo2018pathint}).


\begin{acknowledgments}
We acknowledge financial support by MUR (Ministero dell’Universit{\`a} e della Ricerca) through the PNRR MUR project PE0000023-NQSTI.
    We thank J{\"u}rgen Stockburger for useful discussions.
\end{acknowledgments}


\appendix
\clearpage
\begin{widetext}

\section{\label{app:gaussian_physical}
    Gaussian evolution in physical time
}

In Eq.~\eqref{eq:gaussian_exact} we provided an exact expression that describes the dynamics of an open quantum system in contact with a Gaussian environment in terms of the Keldysh contour.
In this appendix we write Eq.~\eqref{eq:gaussian_exact} in physical-time representation, showing that the resulting expression is the one that has been previously reported in Ref.~\cite{diosi2014gaussian}.

We do that by splitting $\iint_{\gamma(t)}dz_1 dz_2$ into four terms, dividing $\gamma(t)$ in its constituting branches $\gamma_-(t)$ and $\gamma_+(t)$.
For example, the term with $z_1,z_2 \in \gamma_-(t)$ can be treated as
\begin{align}
    -\frac{1}{2} \sum_{\alpha,\beta} \int_0^t d\tau_1 d\tau_2 & C_{\alpha,\beta}(\tau_{1-}, \tau_{2-}) A_\alpha(\tau_{1-}) A_\beta(\tau_{2-}) \notag \\
    &= -\frac{1}{2} \sum_{\alpha,\beta} \iint_0^t d\tau_1 d\tau_2 \bigg[ \theta(\tau_1 - \tau_2) c_{\alpha\beta}(\tau_1,\tau_2) + \theta(\tau_2 - \tau_1) c_{\beta\alpha}(\tau_2, \tau_1) \bigg] A_\alpha(\tau_{1-}) A_\beta(\tau_{2-}) \notag \\
    &= -\frac{1}{2} \sum_{\alpha,\beta} \iint_0^t d\tau_1 d\tau_2  \theta(\tau_1 - \tau_2) c_{\alpha\beta}(\tau_1,\tau_2) \qty{ A_\alpha(\tau_{1-}), A_\beta(\tau_{2-}) },
\end{align}
where in the last equality we performed the substitution $\alpha \leftrightarrow \beta$ and $\tau_1 \leftrightarrow \tau_2$.
Similarly, we can treat the other cases, to obtain
\begin{equation} \label{eq:ferialdistep}
    \frac{1}{2} \sum_{\alpha,\beta} \iint_0^t d\tau_1 d\tau_2 \, c_{\alpha\beta}(\tau_1,\tau_2) \bigg[ \{ A_\alpha(\tau_{1+}), A_\beta(\tau_{2-}) \}
    - \theta(\tau_1 - \tau_2) \{ A_\alpha(\tau_{1-}), A_\beta(\tau_{2-}) \}
    - \theta(\tau_2 - \tau_1) \{ A_\alpha(\tau_{1+}), A_\beta(\tau_{2+}) \} \bigg]
\end{equation}
for the exponent in Eq.~\eqref{eq:gaussian_exact}.
Finally, we can turn the contour ordering into a standard time ordering with $\varrho(0)$ outside of it, provided we prescribe how the $A_\alpha(\tau)$ operators should be placed with respect to $\varrho(0)$.
To do that, we use the left-right notation $A_\alpha^L(\tau) X \coloneqq A_\alpha(\tau) X$, $A_\alpha^R(\tau) X \coloneqq X A_\alpha(\tau)$.
Remembering Fig.~\ref{fig:keldysh} and using the fact that inside the contour ordering operators commute, we can therefore perform the following substitutions in Eq.~\eqref{eq:ferialdistep}:
\begin{subequations}
    \begin{gather}
        \{ A_\alpha(\tau_{1+}), A_\beta(\tau_{2-}) \} \mapsto 2 A_\beta^L(\tau_2) A_\alpha^R(\tau_1), \\
        \{ A_\alpha(\tau_{1-}), A_\beta(\tau_{2-}) \} \mapsto 2 A_\alpha^L(\tau_1) A_\beta^L(\tau_2), \\
        \{ A_\alpha(\tau_{1+}), A_\beta(\tau_{2+}) \} \mapsto 2 A_\beta^R(\tau_2) A_\alpha^R(\tau_1).
    \end{gather}
\end{subequations}
The final result is
\begin{equation} \label{eq:Ferialdi}
    \varrho(t) = \mathcal{T} \exp \bigg\{ \sum_{\alpha,\beta} \int_0^t d\tau_1 d\tau_2 \, c_{\alpha\beta}(\tau_1,\tau_2)
    \bigg[ A_\beta^L(\tau_2) A_\alpha^R(\tau_1) - \theta(\tau_1 - \tau_2) A_\alpha^L(\tau_1) A_\beta^L(\tau_2) - \theta(\tau_2 - \tau_1) A_\beta^R(\tau_2) A_\alpha^R(\tau_1) \bigg] \bigg\} \varrho(0),
\end{equation}
which is precisely the expression reported in Ref.~\cite{diosi2014gaussian}.


\section{\label{app:initial_corr}
    Initial system-environment correlations
}

In this appendix we prove Eq.~\eqref{eq:exp_corr}, which shows how the presence of initial correlations between system and environment in the form~\eqref{eq:initial_canonical} can be dealt with in our framework.

Let us first introduce the following propagator on the modified Keldysh contour $\gamma(t,b)$:
\begin{equation}
    W(z_1, z_2) \coloneqq \begin{cases}
        \cev{\mathcal T} \exp[-i \int_{z_2}^{z_1} H(z) dz], & z_1 \succ z_2, \\
        \vec{\mathcal T} \exp[i \int_{z_1}^{z_2} H(z) dz], & z_2 \succ z_1,
    \end{cases}
\end{equation}
where $H(z) \coloneqq H_0(z) + V$ and where $H_0(z)$ is defined in Eq.~\eqref{eq:H0_mod}.
This operator is not unitary unless both its arguments lie on the horizontal branches, but familiar properties still hold~\cite{stefanucci2013book}:
\begin{equation}
    W(z,z) = \mathbbm{1},
    \qquad
    W(z_1,z_2) = W(z_1,z_3) W(z_3,z_2),
    \qquad
    \frac{d}{dz_1} W(z_1,z_2) = \begin{cases}
        -i H(z_1) W(z_1, z_2), & z_1 \succ z_2, \\
        i W(z_1,z_2) H(z_2), & z_2 \succ z_1.
    \end{cases}
\end{equation}
We also define $W_0(z_1,z_2)$ similarly to $W(z_1,z_2)$ but with $H_0(z)$ instead of $H(z)$, as reported in Eq.~\eqref{eq:W0}.

Now, we clearly have from Eq.~\eqref{eq:initial_canonical} that
\begin{equation}
    \rho_{SE}(0) = \frac{1}{Z} W(-ib, ib)
    = \frac{1}{Z} W(-ib, 0) W(0,ib),
\end{equation}
and therefore, propagating at time $t$,
\begin{equation}
    \rho_{SE}(t) = U(t,0) \rho_{SE}(0) U(0,t)
    = W(t-ib, -ib) \rho_{SE}(0) W(ib, t+ib)
    = \frac{1}{Z} W(t-ib,0) W(0,t+ib).
\end{equation}
Now we move to the interaction picture,
\begin{align}
    \varrho_{SE}(t) &= e^{iH_0 t} \rho_{SE}(t) e^{-iH_0 t} = W_0(-ib, t-ib) \rho_{SE}(t) W_0(t+ib, ib) \notag \\
    &= \frac{1}{Z} \qty[ W_0(-ib, t-ib) W(t-ib,0) ] \qty[ W(0,t+ib) W_0(t+ib, ib) ].
\end{align}
At this point we need the following identity.
Given points $z_1 \succ z_2$ and $z_1 \succ \overline{z}$,
\begin{subequations}
    \begin{gather}
        W(z_1,z_2) = W_0(z_1, \overline{z}) W_I(z_1,z_2; \overline{z}) W_0(\overline{z},z_2), \\
        W_I(z_1,z_2; \overline{z}) \coloneqq \cev{\mathcal T} \exp[ -i \int_{z_2}^{z_1} W_0(\overline{z},z) V(z) W_0(z, \overline{z})dz ].
    \end{gather}
\end{subequations}
This can be proved by showing that both sides of the equation solve the same differential equation.
For the left-hand side, we know that
\begin{equation}
    \frac{d}{dz_1} W(z_1,z_2) = -iH(z_1)W(z_1,z_2).
\end{equation}
For the right-hand side we have
\begin{align}
    & \frac{d}{dz_1} \qty[ W_0(z_1, \overline{z}) W_I(z_1,z_2; \overline{z}) W_0(\overline{z},z_2) ] \notag \\
    &= -i H_0 W_0(z_1, \overline{z}) W_I(z_1,z_2;\overline{z}) W_0(\overline{z},z_2) - i W_0(z_1,\overline{z}) W_0(\overline{z},z_1) V(z_1) W_0(z_1, \overline{z}) W_I(z_1,z_2;\overline{z}) W_0(\overline{z},z_2) \notag \\
    &= -i H(z_1) W_0(z_1, \overline{z}) W_I(z_1,z_2;\overline{z}) W_0(\overline{z},z_2),
\end{align}
where we used the fact that
\begin{equation}
    \frac{d}{dz_1} W_I(z_1,z_2; \overline{z}) = -i \qty[ W_0(\overline{z},z_1) V(z_1) W_0(z_1, \overline{z}) ] W_I(z_1,z_2;\overline{z}).
\end{equation}

We can then write
\begin{subequations}
    \begin{gather}
        W_0(-ib, t-ib) W(t-ib,0) = W_I(t-ib, 0; -ib) W_0(-ib, 0), \\
        W(0,t+ib) W_0(t+ib, ib) = W_0(0,ib) W_I(0, t+ib; ib).
    \end{gather}
\end{subequations}
As a consequence,
\begin{align}
    \varrho_{SE}(t) &= \frac{1}{Z} W_I(t-ib, 0; -ib) W_0(-ib,0) W_0(0,ib) W_I(0, t+ib; ib) \notag \\
    &= \frac{1}{Z} W_I(t-ib, 0; -ib) W_0(-ib, ib) W_I(0, t+ib; ib) \notag \\
    &= \cev{\mathcal T} \exp[ -i \int_{0}^{t-ib} W_0(-ib,z) V(z) W_0(z,-ib)dz ] \frac{e^{-2b H'_0}}{Z} \cev{\mathcal T} \exp[ -i \int_{t+ib}^0 W_0(ib,z) V(z) W_0(z,ib) dz ],
\end{align}
which can also be arranged as in Eq.~\eqref{eq:exp_corr}.


\section{Contour ordered exponentials and energy statistics}
\label{sec:therm}

Our goal is to write an expression of the form \ref{eq:R_solution} for the MGF \ref{eq:MGF}.
We start by expressing the exponentials in $\lambda$ as integrations over a dummy variable $\zeta$
\begin{align}  \notag
  M_{\Lambda}(t, \lambda) & =  \Tr\bigg[ \cev{\mathcal{T}} \exp\bigg(- i \int_{\lambda}^{t + \lambda} H(\tau) d \tau \bigg)
    \exp\bigg(- i \int_{0}^{\lambda} \Lambda(0) d \zeta \bigg)  \rho_{SE}(0)
    \\ &  \times \cev{\mathcal{T}}   \exp \bigg(- i \int_{t}^{0} H(\tau) d \tau \bigg)   \exp \bigg(- i \int_{t+\lambda}^{t} \Lambda(\tau) d \zeta \bigg)
      \bigg].
      \end{align}
The product of the first two exponential operators can be seen as a single ordered product on a domain going from $0$ to $t+\lambda$. By analogy with the case treated in section. \ref{sec:keldysh}, we call $\gamma_-'(\lambda)$ the part of this domain going from $0$ to $\lambda$ and $\gamma_-(t,\lambda)$ the remaining part, going from $\lambda$ to $t + \lambda$.
With the same reasoning, the exponential operators on the right can be written as a single ordered product on a backward track $ \gamma_+(t) \oplus \gamma_+'(t, \lambda) $ so that we can write (see also Fig. \ref{fig:keldysh3} in the main text)
    \begin{align} \label{eq:MGFman1}
  M_{\Lambda}(t,\lambda) =  \Tr\bigg[\cev{\mathcal{T}}\exp\bigg(- i \int_{\gamma_-'(\lambda) \oplus \gamma_-(t, \lambda)} H(z) d z \bigg)
   \rho_{SE}(0)   \cev{\mathcal{T}} \exp\bigg(- i \int_{\gamma_+(t) \oplus \gamma_+'(t, \lambda)} H(z) d z \bigg)
      \bigg],
\end{align}
where $H(z) = H_{+}(\tau)$ for $z = \tau \in \gamma_+(t) \oplus \gamma_+'(t, \lambda)$ and $H(z) = H_{-}(\tau)$ for $z = \tau \in\gamma_-'(\lambda) \oplus \gamma_-(t, \lambda)$ with
\begin{align}
    H_{-}(\tau) \coloneqq \begin{cases}
     \Lambda(0)   \: \quad \quad \quad  \tau \leq \lambda,  \\
       H(\tau - \lambda) \quad  \lambda \leq \tau \leq \lambda + t,
    \end{cases}
  \quad \quad
    H_{+}(\tau) \coloneqq
      \begin{cases}
      H(\tau) \quad  \quad \quad \tau \leq t,
        \\
        \Lambda(t)  \, \quad \quad \quad  t \leq \tau \leq t+\lambda.
    \end{cases} \end{align}
For ease of notation we can distinguish the integral on the forward/backward branches by adding a sign $\pm$ in the integration extrema ($-$ for the forward branch and $+$ for the backward one):
    \begin{align} \label{eq:exppm}
  M_{\Lambda}(t,\lambda) =  \Tr\bigg[\cev{\mathcal{T}}\exp\bigg(- i \int_{0}^{(t + \lambda)_-} H(z) d z \bigg)
   \rho_{SE}(0)
     \cev{\mathcal{T}}\exp\bigg(- i \int_{(t+\lambda)_+}^{0} H(z) d z \bigg)
      \bigg].
\end{align}
As done in the case of the standard time evolution we can compact the trace in a single time-ordered expression by introducing the contour $\gamma(t,\lambda)$ in Fig. \ref{fig:keldysh3}:
\begin{equation}
    M_{\Lambda}(t,\lambda) = \Tr \bigg[\cev{\mathcal{T}}  \exp \bigg(
 -i\int_{\gamma(t,\lambda)} H(z) dz \bigg) \rho_{SE}(0)  \bigg].
\end{equation}
Since we are dealing with the open system case is convenient to introduce the interaction picture. For two generic operators $H_0(z)$ and $V(z)$ with $z \in \gamma(t,\lambda)$ and such that
$H(z) = H_0(z) + V(z)$,
we can define
\begin{align} \label{eq:Wnew}
    W(z_1, z_2) = \begin{cases}
       \cev{\mathcal{T}} \big\{ \exp\big(- i \int_{z_2}^{z_1} H(z) dz\big) \big\} & z_2 \geq z_1, \\
     \vec{\mathcal{T}} \big\{ \exp\big( i \int_{z_1}^{z_2} H(z) dz\big)\big\} & z_2 < z_1 .
    \end{cases}
\end{align}
In this way, we will obtain
\begin{align} \label{eq:Wpm2}
    W(\tau_-, 0) = W_0(\tau_-,0) W_I(\tau_-,0), \quad  \quad
   W(0, \tau_+) =  W_I(0,\tau_+) W_0(0, \tau_+),
\end{align}
where $W_I(z_1,z_2)$ satisfies the definition in Eq.~\eqref{eq:Wnew} with $H(z)$ replaced by
$   \tilde{V}(z) = W_0(0,z) V(z) W_0(z,0)$.
Combining Eqs. \eqref{eq:exppm}, \eqref{eq:Wnew} and \eqref{eq:Wpm2}  we can express the generating function as
\begin{align} \notag
M_{\Lambda}(t,\lambda) & = \Tr \big[W[(t+\lambda)_-, 0] \rho_{SE}(0) W[0, (t+\lambda)_+] \big]
\\
& =
 \Tr \big[ W_0[(t+ \lambda)_-, 0] W_I[(t+\lambda)_-,0] \rho_{SE}(0) W_I[0, (t+\lambda)_+]
W_0[0, (t+\lambda)_+] \big].
\end{align}
To simplify the expression above, let us work under the assumption that $H_0(\tau)$ commutes with $\Lambda(0), \Lambda(t)$ for all values of $\tau$ and that $\Lambda(t) = \Lambda(0)$. The latter condition is satisfied, for instance, when computing the statistics of heat using a TPEM or when computing the statistics of work in a cyclic process. In this case we have
\begin{align}
    W_0\big[0, (t+\lambda)_+\big] W_0\big[(t+\lambda)_-,0\big]
    & = \cev{\mathcal{T}}  \exp\bigg(- i \int_{(t+\lambda)_+}^0 H_0(z) dz \bigg) \cev{\mathcal{T}}     \exp \bigg(-i \int_0^{(t+\lambda)_- }  H_0(z) dz\bigg)  = \mathds{1}.
\end{align}
This leads to

\begin{align}
    M_{\Lambda}(t,\lambda) & = \Tr[W_I((t+\lambda)_-,0 ) \rho_{SE}(0) W_I(0, (t+\lambda)_+) ]
    \\     & = Tr\bigg[\cev{\mathcal{T}} \exp \Big(- i \int_0^{(t+\lambda)_- } \tilde{V}(z) dz \Big) \rho_{SE}(0)\cev{\mathcal{T}} \exp \Big(- i \int_{(t+\lambda)_+}^{0} \tilde{V}(z)dz\Big) \bigg]
        \\
    & = Tr\bigg[\cev{\mathcal{T}} \exp\Big( - i\int_{\gamma(t,\lambda)} \tilde{V}(z) dz\Big)
 \rho_{SE}(0) \bigg].
\end{align}
If we are computing the statistics of heat the measurement operators $\Lambda(0)$ and $\Lambda(t)$ coincide with the environmental Hamiltonian, i.e. $\Lambda(0) = \Lambda(t) = H_E$.
We assume the physical Hamiltonian to be given by Eq. \eqref{eq:totH}, but we include the possibility of an explicit time dependence in $H_S$

\begin{subequations} \label{eq:totHappapp}
    \begin{gather}
        H(t) = H_0(t) + V, \\
        H_0(t) = H_S(t) \otimes \mathbbm{1}_E + \mathbbm{1}_S \otimes H_E, \label{eq:H0app} \\
        V = \sum_\alpha A_\alpha \otimes B_\alpha.
        \label{eq:interactionapp}
    \end{gather}
\end{subequations}
 Under these assumptions $V(z) =0$ on $\gamma_-'(\lambda)$ and $\gamma_+'(t,\lambda)$ while on $\gamma_-(t,\lambda)$ and $\gamma_+(t)$ we have

\begin{equation}
    V[(\tau + \lambda)_-] = V(\tau_+) = V = \sum_{\alpha} A_{\alpha} B_{\alpha}.
\end{equation}
Hence, in the interaction picture we obtain, in the forward branch:
\begin{align}
    \tilde{V}[(\tau + \lambda)_-] & = W_0(0, (\tau+\lambda)_-) V W_0((\tau+\lambda)_-,0)
     = e^{i \lambda H_E} e^{i \tau H_0} \sum_{\alpha} A_{\alpha} B_{\alpha}
    e^{-i \tau H_E} e^{- i \lambda H_E}
    \\
    & = \sum_{\alpha} e^{i \tau H_S} A_{\alpha} e^{- i \tau H_S} \otimes e^{i (\tau+ \lambda) H_E} B_{\alpha} e^{- i (\tau+ \lambda) H_E}
    = \sum_{\alpha} A_{\alpha}(\tau) \otimes B_{\alpha}(\tau +\lambda),
\end{align}
while in the backward branch we obtain:

\begin{equation}
    \tilde{V}(\tau_+) = W_0(0,\tau_+) V W_0(\tau_+,0) = \sum_{\alpha} A_{\alpha}(\tau) \otimes B_{\alpha}(\tau).
\end{equation}
This leads to the equation $M_{H_E} = \Tr[\rho_{SE}(t,\lambda)]$ with
\begin{equation} \label{eq:ordmodV}
    \rho_{SE}(t,\lambda) = \cev{\mathcal{T}}\exp\bigg(-i \int_{\gamma(t)} \mathcal{V}_{\lambda}(z) dz\bigg) \rho_{SE}(0) .
\end{equation}
where we introduced the modified interaction operator
\begin{align}
    \mathcal{V}_{\lambda}(z) = \begin{cases}
    \sum_{\alpha} A_{\alpha}(\tau) \otimes B_{\alpha}(\tau+\lambda) &  z= \tau_-,\\
     \sum_{\alpha} A_{\alpha}(\tau) \otimes B_{\alpha}(\tau)  &  z= \tau_+.
    \end{cases}
\end{align}
Notice that the formula \eqref{eq:ordmodV} the standard contour $\gamma(t)$ (the one in Fig. \ref{fig:keldysh}) appears and not the modified contour $\gamma(t,\lambda)$ (in Fig. \ref{fig:keldysh3}).
Now we want to trace away the degrees of freedom of the environment, and obtain an equation for the reduced state
\begin{equation} \label{eq:reduced}
    \rho(t,\lambda) = \Tr_E[\rho_{SE}(t,\lambda)].
\end{equation}
In analogy with the treatment of sec. \ref{sec:gaussian}, if the environment is Gaussian and the $B$ operators are linear in the environmental creation/annihilation modes, we can use the Wick theorem that will allow us to write the environment averages in the time-ordered exponential as linear combinations of products of correlation functions of the form

\begin{equation}
    C_{\alpha, \beta}(z_1,z_2) = \Tr[\cev{\mathcal{T}}\big\{B_{\alpha}(z_1) B_{\beta} (z_2)\big\} \Omega],
\end{equation}
where now $z$ lives on the modified contour $\gamma(t,\lambda)$.
In analogy to what we did in Sec. \ref{sec:gaussian}, the trace in Eq. \eqref{eq:reduced}, after replacing inside Eq. \eqref{eq:ordmodV}, can be simplified by using the Wick theorem.
After introducing a set of noises $\xi_{\alpha}(z)$ with the correct correlation function, i.e. such that
\begin{align}
    \langle \xi_{\alpha}(z_1)  \xi_{\beta}(z_2) \rangle = C_{\alpha \beta}(z_1,z_2),
\end{align}
we follow the steps done in Sec. \ref{sec:gaussian} and we compute the values in the forward and backward branches

\begin{align}
C_{\alpha \beta}(\tau_{1+}, \tau_{2+}) = \theta(\tau_1-\tau_2) c_{\beta \alpha}(\tau_2,\tau_1) + \theta(\tau_2- \tau_1) c_{\alpha \beta}(\tau_1, \tau_2), \\
C_{\alpha \beta}(\tau_{1-}, \tau_{2-}) = \theta(\tau_1-\tau_2) c_{\alpha \beta}^{(\lambda \lambda)}(\tau_2,\tau_1) + \theta(\tau_2- \tau_1) c^{(\lambda\lambda)}_{\alpha \beta}(\tau_1, \tau_2), \\
  C_{\alpha \beta }(\tau_{1+}, \tau_{2-}) = c_{\alpha \beta}^{(\lambda)}(\tau_1,\tau_2), \quad \quad
    C_{\alpha \beta}(\tau_{1-}, \tau_{2+})= c_{\alpha \beta}^{(\lambda)}(\tau_2,\tau_1)
\end{align}
where we introduced
\begin{align}
   c_{\alpha\beta}^{(\lambda\lambda)}(\tau_1,\tau_2) & = \Tr [ B_{\alpha}(\tau_1 +\lambda) B_{\beta}(\tau_2 +\lambda) \Omega] = c_{\alpha\beta} (\tau_1,\tau_2), \\
c_{\alpha \beta}(\tau_1,\tau_2) & = \Tr [B_{\alpha}(\tau_1) B_{\beta}(\tau_2) \Omega],
\\
 c_{\alpha\beta}^{(\lambda)}(\tau_1,\tau_2) & = \Tr[B_{\alpha}(\tau_1) B_{\beta}(\tau_2+\lambda) \Omega].
\end{align}
In conclusion, if we define the Keldysh-rotated versions of the forward and backward components of the noise $\xi(z)$, i.e.
\begin{align}
    \nu_{\alpha}(t) = \frac{1}{2}[\xi_{\alpha}^-(t) + \xi_{\alpha}^+(t)], \qquad
    \eta_{\alpha}(t) = \frac{1}{2}[\xi_{\alpha}^-(t) - \xi_{\alpha}^+(t)]
\end{align}
we obtain the following correlation functions
\begin{align} \notag
    \langle \nu_{\alpha}(t) \nu_{\beta}(s) \rangle
&  = \frac{1}{4} \langle [\xi_{\alpha}^-(t) + \xi_{\alpha}^+(t)] [\xi_{\beta}^-(s) + \xi_{\beta}^+(s)] \rangle   =
 \frac{1}{4} \langle \xi_{\alpha}^-(t)  \xi_{\beta}^-(s)
 + \xi_{\alpha}^-(t) \rangle \xi_{\beta}^+(s) + \xi_{\alpha}^+(t) \rangle \xi_{\beta}^-(s)
 + \xi_{\alpha}^+(t) \rangle \xi_{\beta}^+(s) \rangle
 \\ \notag
 & = \frac{1}{4} [ \theta(t-s) c_{\alpha \beta}(t,s) + \theta(s-t) c_{\beta \alpha}(s,t)
  + c_{\alpha \beta}^{(\lambda)}(t,s)  + c_{\beta \alpha}^{(\lambda)}(s,t) +
  \theta(s-t) c_{\beta \alpha}(s,t) + \theta(t-s) c_{\alpha \beta}(t,s)]
 \\ &  = \frac{1}{2} \Re c_{\alpha \beta}(t,s) + \frac{1}{4}[c_{\alpha \beta}(t, s+ \lambda) +
  c_{\alpha \beta}^*(t,s-\lambda)],
\end{align}

\begin{align} \notag
    \langle \eta_{\alpha}(t) \eta_{\beta}(s) \rangle
 & = \frac{1}{4} \langle [\xi_{\alpha}^-(t) - \xi_{\alpha}^+(t)] [\xi_{\beta}^-(s) - \xi_{\beta}^+(s)] \rangle   =
 \frac{1}{4} \langle \xi_{\alpha}^-(t)  \xi_{\beta}^-(s)
 + \xi_{\alpha}^-(t) \rangle \xi_{\beta}^+(s) - \xi_{\alpha}^+(t) \rangle \xi_{\beta}^-(s)
 - \xi_{\alpha}^+(t) \rangle \xi_{\beta}^+(s) \rangle
 \\ \notag
 & = \frac{1}{4} [ \theta(t-s) c_{\alpha \beta}(t,s) + \theta(s-t) c_{\beta \alpha}(s,t)
  - c_{\beta \alpha}^{(\lambda)}(s,t)  - c_{\alpha \beta}^{(\lambda)}(t,s) + \theta(t-s) c_{\beta \alpha}(s,t) + \theta(s-t) c_{\alpha \beta}(t,s)]
 \\ & = \frac{1}{2} \Re c_{\alpha \beta}(t,s) - \frac{1}{4}[c_{\alpha \beta}(t, s+ \lambda) +
  c_{\alpha \beta}^*(t,s-\lambda)],
\end{align}

\begin{align} \notag
    \langle \nu_{\alpha}(t) \eta_{\beta}(s) \rangle
 & = \frac{1}{4} \langle [\xi_{\alpha}^-(t) + \xi_{\alpha}^+(t)] [\xi_{\beta}^-(s) - \xi_{\beta}^+(s)] \rangle   =
 \frac{1}{4} \langle \xi_{\alpha}^-(t)  \xi_{\beta}^-(s)
 - \xi_{\alpha}^-(t) \rangle \xi_{\beta}^+(s) + \xi_{\alpha}^+(t) \rangle \xi_{\beta}^-(s)
 - \xi_{\alpha}^+(t) \rangle \xi_{\beta}^+(s) \rangle
 \\ \notag
 & = \frac{1}{4} [ \theta(t-s) c_{\alpha \beta}(t,s) + \theta(s-t) c_{\beta \alpha}(s,t)
  - c_{\beta \alpha}^{(\lambda)}(s,t)  + c_{\alpha \beta}^{(\lambda)}(t,s) - \theta(t-s) c_{\beta \alpha}(s,t) - \theta(s-t) c_{\alpha \beta}(t,s)]
\\ &  = \frac{1}{2} {\rm sign}(t-s) \Im c_{\alpha \beta}(t,s) + \frac{1}{4}[c_{\alpha \beta}(t, s+ \lambda) -
  c_{\alpha \beta}^*(t,s-\lambda)].
\end{align}

\section{\label{app:partial_wick}
    Partial scalar Wick's theorem
}

In this appendix we show how Eq.~\eqref{eq:partial_isserlis} can be established.
As we already said in Sec.~\ref{sec:reduction}, since $\nu_j$ and $\eta_j$ are correlated Gaussian noises, we can use the scalar Wick's theorem, which we repeat here for the reader's convenience:
\begin{equation} \label{eq:isserlis_app}
    \mathbb{E}[ \chi_1 \ldots \chi_{2m} ] = \frac{1}{m! 2^m} \sum_{\sigma \in \mathfrak{S}_{2m}} \prod_{j=1}^m \mathbb{E}[\chi_{\sigma(2j-1)} \chi_{\sigma(2j)}],
\end{equation}
where $\{\chi_j\}_{j=1}^{2m}$ is a collection of generic Gaussian noises.
Let us consider the case in which $\chi_j = \nu_j$ for $j=1,\ldots,k$ and $\chi_j = \eta_j$ for $j=k+1,\ldots,2m$, where $k \geq m$.
Since $\mathbb{E}[\eta_i \eta_j] = 0$ [see Eq.~\eqref{eq:rot_corr}], in Eq.~\eqref{eq:isserlis_app} we can restrict to those permutations that couple each $\eta_j$ with a $\nu_j$:
\begin{multline}
    \mathbb{E}[\nu_1 \ldots \nu_k \eta_{k+1} \ldots \eta_{2m}] = \frac{\theta(k-m)}{(k-m)!2^{k-m}} \\
    \times \sum_{\sigma \in \mathfrak{S}_k} \mathbb{E}[\nu_{\sigma(1)} \eta_{k+1}] \ldots \mathbb{E}[\nu_{\sigma(2m-k)} \eta_{2m}]
    \mathbb{E}[\nu_{\sigma(2m-k+1)} \nu_{\sigma(2m-k+2)}] \ldots \mathbb{E}[\nu_{\sigma(k-1)} \nu_{\sigma(k)}].
\end{multline}
If we insert this expression into Eq.~\eqref{eq:gigante}, we see that, because of the integration structure and the fact that operators commute inside the $\cev{\mathcal{T}}$ sign, each permutation contributes equally.
As a consequence, we can make the substitution
\begin{equation}
    \mathbb{E}[\nu_1 \ldots \nu_k \eta_{k+1} \ldots \eta_{2m}] \mapsto \frac{\theta(k-m) k!}{(k-m)! 2^{k-m}}
    \mathbb{E}[\nu_1 \eta_{k+1}] \ldots \mathbb{E}[\nu_{2m-k} \eta_{2m}] \mathbb{E}[\nu_{2m-k+1} \nu_{2m-k+2}] \ldots \mathbb{E}[\nu_{k-1} \nu_k].
\end{equation}
At this point we want to use Wick's theorem again to regroup the expectation values involving $\nu_j$ only into a single expectation value.
To do that, we use again that Eq.~\eqref{eq:gigante} is invariant under permutations of the indexes of the $\nu_j$:
\begin{align}
    \mathbb{E}[\nu_{2m-k+1} \nu_{2m-k+2}] \ldots \mathbb{E}[\nu_{k-1} \nu_k] &\mapsto \frac{1}{(2k-2m)!}
    \sum_{\pi \in \mathfrak{S}_{2m-k+1}^k} \mathbb{E}[\nu_{\pi(2m-k+1)} \nu_{\pi(2m-k+2)}] \ldots \mathbb{E}[\nu_{\pi(k-1)} \nu_{\pi(k)}] \notag \\
    &= \frac{(k-m)! 2^{k-m}}{(2k-2m)!} \mathbb{E}[\nu_{2m-k+1} \ldots \nu_k],
\end{align}
where $\mathfrak{S}_a^b$ is the set of permutations of $\{a, \ldots, b\}$, for $a \leq b$.
We conclude that
\begin{equation}
    \mathbb{E}[\nu_1 \ldots \nu_k \eta_{k+1} \ldots \eta_{2m}] \mapsto \frac{\theta(k-m) k!}{(2k-2m)!}
    \mathbb{E}[\nu_1 \eta_{k+1}] \ldots \mathbb{E}[\nu_{2m-k} \eta_{2m}] \mathbb{E}[\nu_{2m-k+1} \ldots \nu_k],
\end{equation}
which is precisely Eq.~\eqref{eq:partial_isserlis} if we define $G_{a,b} \coloneqq -i \mathbb{E}[\nu_a \eta_b]$.


\section{\label{app:wigner}
    Gaussian measurements with Wigner transforms
}

In Sec.~\ref{sec:suffmeas} we discussed sufficient conditions under which the stochastic von Neumann equation~\eqref{eq:stocvon} can be obtained in a scheme in which the noise $\nu_{\alpha}(t)$ is described in terms of a one-shot environment measurement performed at the beginning of the dynamics.
Such conditions, reported in Eq.~\eqref{eq:wigner_conditions}, can be established using the Wigner function formalism~\cite{polkovnikov2010phase}, as detailed below.

Let us first pretend that the environment is composed by a single mode with associated position and momentum operators $X$ and $P$.
Thanks to the independence of the modes, it is straightforward to generalize what we will say to the multimode scenario.
Moreover, given an operator $O(X,P)$, we follow Ref.~\cite{polkovnikov2010phase} and define its Wigner transform as the following scalar function of position and momentum variables $x$ and $p$:
\begin{equation}
    O_W(x,p) \coloneqq \int_{-\infty}^\infty d\xi \mel{x-\frac{\xi}{2}}{O}{x+\frac{\xi}{2}} e^{i p\xi}.
\end{equation}

Now, given the initial environment state $\Omega$, consider the non-normalized product [see Eq.~\eqref{eq:omega_y}]
\begin{equation} \label{eq:omega_x_tilde}
    \widetilde{\Omega}_{x'} \coloneqq M_{X,x'} \Omega M^\dagger_{X,x'}
\end{equation}
that one should consider when performing a measurement of the position $X$, obtaining the result $x'$.
Using the fact that
\begin{equation}
    M_{X,x'} \ket{x} = \frac{1}{\qty(2\pi \sigma_X^2)^{1/4}} \exp[ -\frac{(x-x')^2}{4\sigma_X^2} ] \ket{x},
\end{equation}
as follows from Eq.~\eqref{eq:M_Y}, we can easily write the Wigner transform of Eq.~\eqref{eq:omega_x_tilde}:
\begin{align}
    \widetilde{\Omega}_{x',W}(x,p) &= \frac{1}{\sqrt{2\pi\sigma_X^2}} \int_{-\infty}^\infty d\xi \mel{x-\frac{\xi}{2}}{\Omega}{x+\frac{\xi}{2}} e^{ip\xi} \exp[ -\frac{(x-\frac{\xi}{2}-x')^2}{4\sigma_X^2} ] \exp[ -\frac{(x+\frac{\xi}{2}-x')^2}{4\sigma_X^2} ] \notag \\
    &= \frac{1}{\sqrt{2\pi\sigma_X^2}} \int_{-\infty}^\infty d\xi \mel{x-\frac{\xi}{2}}{\Omega}{x+\frac{\xi}{2}} e^{ip\xi} \exp[ -\frac{(x-x')^2}{2\sigma_X^2} ] \exp[-\frac{\xi^2}{8\sigma_X^2}].
\end{align}
After inserting the following identity
\begin{equation}
   1 = \int_{-\infty}^\infty \delta(\tau-\xi) d\tau = \frac{1}{2\pi} \int_{-\infty}^\infty d\tau \int_{-\infty}^\infty d\epsilon \, e^{i(\tau-\xi)\epsilon},
\end{equation}
replacing the $\xi$ appearing in the last exponential with $\tau$ and doing the integral in $\xi$, we remain with
\begin{equation}
    \widetilde{\Omega}_{x',W}(x,p) = \frac{1}{2\pi} \frac{1}{\sqrt{2\pi\sigma_X^2}} \int_{-\infty}^\infty d\tau \int_{-\infty}^\infty d\epsilon \, \Omega_W(x,p-\epsilon) \exp[ i\tau\epsilon - \frac{\tau^2}{8\sigma_X^2} ] \exp[-\frac{(x-x')^2}{2\sigma_X^2}].
\end{equation}
Since the integral in $\tau$ gives
\begin{equation}
    \int_{-\infty}^\infty d\tau \, \exp[i\tau\epsilon - \frac{\tau^2}{8\sigma_X^2}] = \sqrt{8\pi\sigma_X^2} e^{-2\epsilon^2\sigma_X^2},
\end{equation}
we conclude that the non-normalized post-measured state, conditioned on the outcome $x'$, has a Wigner representation
\begin{equation}
    \widetilde{\Omega}_{x',W}(x,p) = \frac{1}{\pi} \int_{-\infty}^\infty d\epsilon \, \Omega_W(x,p-\epsilon) e^{-2\epsilon^2 \sigma_X^2} \exp[-\frac{(x-x')^2}{2\sigma_X^2}].
\end{equation}
In order to normalize the result [see Eq.~\eqref{eq:omega_y}] we should also calculate
\begin{align}
    \mathbb{P}(x') \coloneqq \Tr \widetilde{\Omega}_{x'} = \frac{1}{2\pi} \iint_{-\infty}^\infty dx \, dp \, \widetilde{\Omega}_{x',W}(x,p)
    = \frac{1}{2\pi \sqrt{2\pi\sigma_X^2}} \iint_{-\infty}^\infty dx \, dp \, \Omega_W(x,p) \exp[-\frac{(x-x')^2}{2\sigma_X^2}],
\end{align}
where we used the change of variable $p \mapsto p-\epsilon$ to eliminate the integral over $\epsilon$.

This tells us that the post-measured state undergoes a perturbation in the momentum when a measurement of the position is performed, whose strength varies according to the measurement precision.
It is worth to notice that if the measurement of the position is extremely imprecise, i.e., $\sigma_X \to \infty$, the post-measured Wigner function remains equal to the original Wigner function.
It is also easy to show that a similar calculation can be done when one performs a measurement of the momentum:
\begin{subequations}
    \begin{gather}
        \widetilde{\Omega}_{p',W}(x,p) = \frac{1}{\pi} \int_{-\infty}^\infty d\epsilon \, \Omega_W(x-\epsilon,p) e^{-2\epsilon^2 \sigma_P^2} \exp[-\frac{(p-p')^2}{2\sigma_P^2}], \\
        \mathbb{P}(p') = \frac{1}{2\pi \sqrt{2\pi\sigma_P^2}} \iint_{-\infty}^\infty dx \, dp \, \Omega_W(x,p) \exp[-\frac{(p-p')^2}{2\sigma_P^2}].
    \end{gather}
\end{subequations}

Combining an imperfect measurement of the position operator with one of the momentum, we obtain the so-called ``heterodyne'' approach.
In this case, having defined $y' = (x',p')$ and assuming the momentum measurement to be performed before the position measurement,
\begin{align}
    \widetilde{\Omega}_{y',W}(x,p) &= \frac{1}{\pi} \int_{-\infty}^\infty d\epsilon_1 \, \widetilde{\Omega}_{p',W}(x,p-\epsilon_1) e^{-2\epsilon_1^2 \sigma_X^2} \exp[-\frac{(x-x')^2}{2\sigma_X^2}] \notag \\
    &= \frac{1}{\pi^2} \iint_{-\infty}^\infty d\epsilon_1 d\epsilon_2 \, \Omega_W(x-\epsilon_2,p-\epsilon_1) e^{-2\epsilon_1^2 \sigma_X^2} e^{-2\epsilon_2^2 \sigma_P^2} \exp[-\frac{(x-x')^2}{2\sigma_X^2}] \exp[-\frac{(p-\epsilon_1-p')^2}{2\sigma_P^2}],
    \label{eq:wigner_omega_y}
\end{align}
where, using the variances in Eq.~\eqref{eq:delta_variances}, we can write for the initial state
\begin{equation}
    \Omega_W(x,p) = \frac{1}{\Delta_X \Delta_P} \exp[-\frac{x^2}{2\Delta_X^2}] \exp[-\frac{p^2}{2\Delta_P^2}].
\end{equation}

Now, suppose we have the conditions (see Eqs. \eqref{eq:wigner_conditions})
\begin{equation}
    \sigma_X \sigma_P \gg 1,
    \qquad
    \sigma_X \ll \Delta_X,
    \qquad
    \sigma_P \ll \Delta_P.
\end{equation}
Then, we also have $\sigma_X \Delta_P \gg 1$ and $\sigma_P \Delta_X \gg 1$.
As a consequence, the integrand in Eq.~\eqref{eq:wigner_omega_y} is strongly peaked around $\epsilon_1 = \epsilon_2 = 0$ and we can approximate the result by removing the $\epsilon$ dependence in $\Omega_W$ and in the exponential containing $p$.
Integrating away the remaining exponential weigths,
\begin{equation}
    \widetilde{\Omega}_{y',W}(x,p) \simeq \frac{1}{2\pi \sigma_X \sigma_P} \Omega_W(x,p) \exp[-\frac{(x-x')^2}{2\sigma_X^2}] \exp[-\frac{(p-p')^2}{2\sigma_P^2}].
\end{equation}
Since $\sigma_X \ll \Delta_X$ and $\sigma_P \ll \Delta_P$, we can also replace the arguments inside $\Omega_W$ with the measurement outcomes, since the exponential factors are then strongly peaked around them:
\begin{equation}
    \widetilde{\Omega}_{y',W}(x,p) \simeq \frac{1}{2\pi \sigma_X \sigma_P} \Omega_W(x',p') \exp[-\frac{(x-x')^2}{2\sigma_X^2}] \exp[-\frac{(p-p')^2}{2\sigma_P^2}].
\end{equation}
Under these assumptions, it is also easy to compute the probability $\mathbb{P}(y')$ by directly integrating the above equation:
\begin{equation} \label{eq:resprob}
    \mathbb{P}(y') = \frac{1}{2\pi} \iint_{-\infty}^\infty dx \, dp \, \widetilde{\Omega}_{y',W}(x,p) \simeq \frac{1}{2\pi} \Omega_W(x',p').
\end{equation}
We conclude that the post-measured state $\Omega_{y'}$ has the Wigner representation
\begin{equation}
    \Omega_{y,W}(x,p) \simeq \frac{1}{\sigma_X \sigma_P} \exp[ -\frac{(x-x')^2}{2\sigma_X^2} ] \exp[-\frac{(p-p')^2}{2\sigma_P^2}].
\end{equation}
This can easily be extended to the multimode scenario as
\begin{equation}
    \Omega_{\mathbf{y}',W}(\mathbf{x},\mathbf{p}) \simeq \frac{1}{\prod_\alpha \sigma_{X_\alpha} \sigma_{P_\alpha}} \exp[ -\sum_\alpha \frac{(x_\alpha - x'_\alpha)^2}{2\sigma^2_{X_\alpha}} ] \exp[ -\sum_\alpha \frac{(p_\alpha - p'_\alpha)^2}{2\sigma^2_{P_\alpha}} ].
\end{equation}

Note that the variances of the post-measured state are given by the inverse of the measurement precision, so that they are much smaller than the variances of the distribution of the outcomes \eqref{eq:resprob}.
We can now use this result to obtain a formula for $E_\alpha^{(\mathbf{y}')}(t)$ in Eq.~\eqref{eq:newavg}.
We first notice that
\begin{equation}
    X_\alpha(t) = X_\alpha \cos(\omega_\alpha t) + \frac{P_\alpha}{m_\alpha \omega_\alpha} \sin(\omega_\alpha t),
\end{equation}
from which we find
\begin{align}
    E_\alpha^{(\mathbf{y}')}(t) &\simeq \frac{1}{2\pi \sigma_{X_\alpha} \sigma_{P_\alpha}} \iint_{-\infty}^\infty dx_\alpha \, dp_\alpha \qty[ x_\alpha \cos(\omega_\alpha t) + \frac{p_\alpha}{m_\alpha \omega_\alpha} \sin(\omega_\alpha t) ] \exp[-\frac{(x_\alpha - x'_\alpha)^2}{2\sigma^2_{X_\alpha}}] \exp[-\frac{(p_\alpha - p'_\alpha)^2}{2\sigma^2_{P_\alpha}}] \notag \\
    &= x'_\alpha \cos(\omega_\alpha t) + \frac{p'_\alpha}{m_\alpha \omega_\alpha} \sin(\omega_\alpha t),
\end{align}
where we used the fact that the Wigner transforms of $X_\alpha$ and $P_\alpha$ are trivially given by $x_\alpha$ and $p_\alpha$, respectively.
This is exactly Eq. \eqref{eq:wigner_result1}.
Eq. \eqref{eq:wigner_result2}
follows from direct calculation starting from \eqref{eq:wigner_result1} and Eq. \eqref{eq:resprob}.
For first, remember that the r.h.s. of Eq. \eqref{eq:wigner_result1} is given by the first of Eqs. \eqref{eq:rot_corr}, that is, Eq. \eqref{eq:anticomcorr}.
More explicitly, this reads (after dropping the index $\alpha$ for ease of notation)

\begin{align} \notag
  & \mathbb{E}_\nu \qty[ \nu(\tau_1) \nu(\tau_2) ]  = \frac{1}{2} Tr[\{X \cos \omega \tau_1 + \frac{P}{m \omega} \sin \omega \tau_1, X \cos \omega \tau_2 + \frac{P}{m \omega} \sin \omega \tau_2 \} \Omega]
   \\ & =
Tr[X^2 \Omega]  \cos \omega \tau_1 \cos \omega \tau_2 +
  \frac{1}{m^2 \omega^2} Tr[P^2  \Omega] \sin \omega \tau_1 \sin \omega \tau_2
  + \frac{1}{2 m \omega} Tr[\{X, P\}  ( \cos \omega \tau_1 \sin \omega \tau_2 + \cos \omega \tau_2 \sin \omega \tau_1) \Omega]. \label{eq:correxpand}
\end{align}

The expectation value of any operator can be written as the integral over the Wigner function of the Wigner trnasform of the operator \cite{polkovnikov2010phase}.
The Weyl symbols of $X^2$ and $P^2$ are the square of the position and momentum arguments of the Wigner function, $x^2$ and $p^2$, respectively.
The Wigner transform of the product is

\begin{align}
\int_{-\infty}^\infty d\xi \mel{x-\frac{\xi}{2}}{XP}{x+\frac{\xi}{2}} e^{i p\xi} = xp + \frac{i \hbar}{2}, \\
\int_{-\infty}^\infty d\xi \mel{x-\frac{\xi}{2}}{PX}{x+\frac{\xi}{2}} e^{i p\xi} = xp - \frac{i \hbar}{2}.
\end{align}

After replacing inside eq. \eqref{eq:correxpand} we have

\begin{align} \notag
   \mathbb{E}_\nu \qty[ \nu(\tau_1) \nu(\tau_2) ]   =
  \frac{1}{2 \pi} \int dx dp  \Omega_{W}(x,p)\big[
  x^2 \cos \omega \tau_1 \cos \omega \tau_2 +  \frac{p^2}{m^2 \omega^2}    \sin \omega \tau_1 \sin \omega \tau_2  +
  \frac{xp}{m \omega} (\cos \omega \tau_1 \sin \omega \tau_2 + \cos \omega \tau_2 \sin \omega \tau_1 ) \big]  .
\end{align}

This coincides with what we get by computing $ \mathbb{E}_y \qty[ E_\alpha^{(\mathbf{y})}(\tau_1) E_\beta^{(\mathbf{y})}(\tau_2) ]  $ with the aid of \eqref{eq:wigner_result1}  and \eqref{eq:resprob}. This concludes the proof of
Eq. \eqref{eq:wigner_result2} for any fixed mode $\alpha$. The generalization to many independent modes is straightforward.

\end{widetext}

\bibliography{bibliography.bib}

\end{document}